\documentclass[twocolumn,nofootinbib,prd]{revtex4}

\usepackage{gabe}

\begin{document}

\title{Homoclinic Orbits around Spinning Black Holes I:\par Exact
Solution for the Kerr Separatrix }

\author{Janna Levin${}^{*,!}$ and Gabe Perez-Giz${}^{**}$}
\email{janna@astro.columbia.edu}
\email{gabe@phys.columbia.edu}
\affiliation{${}^{*}$Department of Physics and Astronomy, Barnard
College of Columbia University, 3009 Broadway, New York, NY 10027 }
\affiliation{${}^{!}$Institute for Strings, Cosmology and Astroparticle
  Physics, Columbia University, New York, NY 10027}
\affiliation{${}^{**}$Physics Department, Columbia University,
New York, NY 10027}


\begin{abstract}
Under the dissipative effects of gravitational radiation, black hole
binaries will transition from an inspiral to a plunge.  The separatrix
between bound and plunging orbits features prominently in the
transition.  For equatorial Kerr orbits, we show that the separatrix
is a homoclinic orbit in one-to-one correspondence with an
energetically-bound, unstable circular orbit.  After providing a
definition of homoclinic orbits, we exploit their correspondence with
circular orbits and derive exact solutions for them.  
This paper focuses on homoclinic behavior in physical space, while
in a companion paper we paint the
complementary phase space portrait. The exact results for the Kerr separatrix
could be useful for analytic or numerical studies of the
transition from inspiral to plunge. 

\end{abstract}

\pacs{04.70.-s, 95.30.Sf, 04.25.-g, 04.20.Jb, 95.10.Ce, 02.30.Ik}

\maketitle

\section{Introduction}

\subsection{Background and Motivation}

A direct observational detection of gravitational waves -- perhaps the
most fundamental prediction of a theory of curved spacetime -- looms
close at hand.  Stellar mass compact objects spiraling into
supermassive black holes have received particular attention as sources
of gravitational radiation for the planned LISA mission
\cite{Glampedakis:2005hs}.
A direct detection of these extreme mass ratio inspirals (EMRIs), as
well as extraction of astrophysics \cite{Flanagan:1997sx1,glampedakis2002:2, drasco2004, drasco2005, drasco2006, lang2006},
requires a thorough knowledge of the underlying dynamics; it is the
motion of the two bodies that shapes the gravitational waveform.  A
well-established approach models the EMRI as an adiabatic progression
through a series of Kerr geodesics \cite{glampedakis2002:2,
Collins:2004ex, drasco2004, drasco2005, drasco2006, Drasco:2006ws,
lang2006}. A transparent depiction of geodesic motion around spinning
black holes is therefore essential, yet seemingly complicated
\cite{chandrasekhar1989,levin2008} and benefits from crucial signposts
in the orbital dynamics.

\begin{figure}[bt]
  \centering
  \begin{minipage}{86mm}
    \includegraphics[width=42mm]{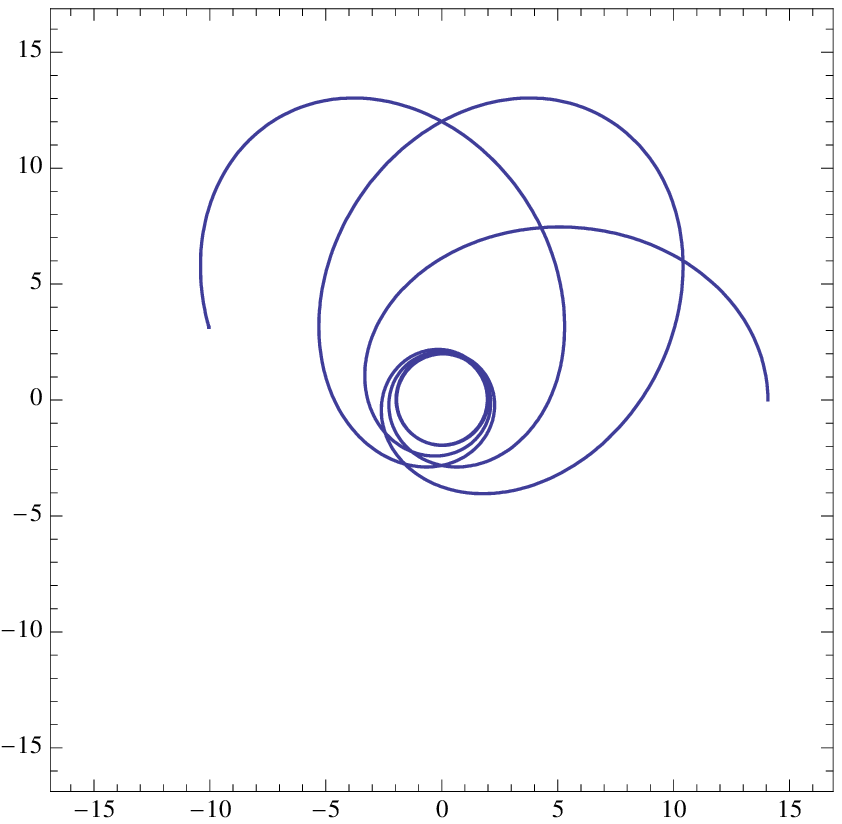}
    \includegraphics[width=42mm]{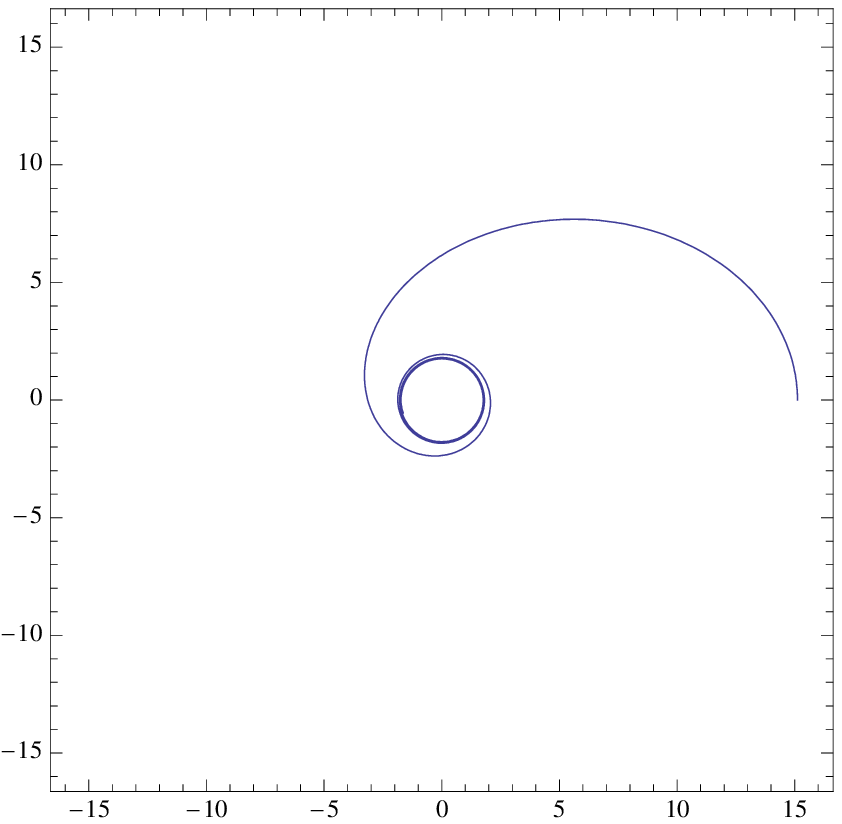}
    \hfill
  \end{minipage}
  \caption{Left: A zoom-whirl orbit. Right: A homoclinic orbit
  approaching an unstable circular orbit.}
  \label{horb}
\end{figure}

We decipher such a crucial signpost here.  In particular, we discuss
an important family of separatrices in Kerr dynamics: the homoclinic
orbits.\footnote{The terms ``homoclinic orbit'' and ``separatrix''
are, in this context, entirely interchangeable, although the former
finds more use in the dynamical systems literature and the latter in
the black hole and gravitational wave literature.}  Around black
holes, the homoclinic orbits are those that asymptotically approach
the same unstable circular orbit in both the infinite future and the
infinite past,\footnote{Orbits that approach two different orbits in
the infinite future and past, in contrast, are called
\emph{heteroclinic orbits}.} as shown on the right of Fig.\
\ref{horb}.
Under the identifier ``separatrix'', homoclinic orbits have already
garnered attention in the black hole literature
\cite{O'Shaughnessy:2002ez,glampedakis2002} -- the homoclinic orbit is
the separatrix between orbits that plunge to the horizon and those
that do not.
The scenario of quasi-circular inspiral through a last stable circular
orbit is a special example of the transition through a zero
eccentricity homoclinic orbit.  Orbits that merge before they have a
chance to circularize will transit through an eccentric homoclinic
orbit of the underlying conservative dynamics.  Any analysis of the
transition from inspiral to plunge will thus run into this special
family.

Homoclinic orbits are also a significant signpost for zoom-whirl
behavior; an extreme form of perihelion precession wherein
trajectories zoom out into quasi-elliptical leaves en route to
apastron and then execute multiple quasi-circular whirls near
periastron before zooming out again, as shown in the left panel of
Fig.\ \ref{horb}.  Though zoom-whirl behavior is sometimes thought to
be associated only with highly eccentric orbits near the separatrix,
we developed a topological criterion for whirliness in
\cite{levin2008} and showed that in the strong-field regime orbits of
any eccentricity can exhibit zoom-whirliness.  Indeed, zoom-whirl
behavior is neither exotic nor rare in the strong field
\cite{levin2008}.  Still, homoclinic orbits are relevant as an
infinite whirl limit in the distribution of geodesics, a connection we
forge in this paper.

Homoclinic orbits are therefore significant in shaping the geography
of black hole orbits. In this first paper in a two part series, we
devote some labor to resolving this landmark in physical space for
equatorial orbits.  (We leave for a future work the non-equatorial
case.)  The pinnacle is an exact solution for equatorial homoclinic
trajectories.  A rarity among relativistic orbits, the exact solution
can make semi-analytic treatment of the eccentric transition to plunge
more wieldy.  In paper II \cite{perez-giz2008}, we describe the
flipside of the coin and detail the phase space portrait of the
homoclinic orbits.  We hope the results will provide cohesion to the
dynamical conversation.

We begin this paper by finding exact expressions for the orbital
parameters of the separatrices and use them to derive Eqs.\
~(\ref{eq:exact}), exact expressions for the trajectories themselves.
As this paper is concerned with the physical space portrait of
homoclinic orbits, we include a final section summarizing the
generality of zoom-whirl behavior and where the separatrix fits in the
spectrum of zoom-whirl orbits.

\subsection{Homoclinic Orbits in the Gravitational Wave Literature}

For context, we note that homoclinic orbits have appeared in the
gravitational wave literature, although not always identified by name.
Ref.\ \cite{O'Shaughnessy:2002ez} analyzed the transition for
equatorial eccentric Kerr orbits using semi-analytic methods.
Gravitational wave snapshots and semi-analytic estimates of the
radiative evolution of orbits near the separatrix appear in
\cite{glampedakis2002}, which also discusses the ``zoom-whirl''
behavior that may be visible during an eccentric transition to plunge.
The discussion of separatrices and their role in eccentric transitions
to plunge is also being discussed for comparable mass systems
\cite{Sperhake:2007gu,levin2008:2}, and an eccentric transition to
plunge, including visible zoom-whirl behavior, has been observed in a
full numerical relativity simulation of the merger of equal mass black
holes \cite{pretorius2007}.

Homoclinic orbits have also been discussed by name in the black hole
literature and are not unique to extreme mass ratio binaries.  The
distinct imprint on a gravitational waveform from the whirl phases or
orbits near the homoclinic set was discussed in \cite{levino2000} for
both Schwarzschild orbits and orbits generated in the Post-Newtonian
(PN) expansion. A program to identify the homoclinic orbits in a
higher-order PN expansion is also underway \cite{levin2008:2}.  Ref.\
\cite{bombelli1992} provides a nice summary of the interesting
phenomenology associated with homoclinic orbits in any dynamical
system, and for the case of Schwarzschild geodesics formally
demonstrates (using a somewhat unphysical example) the onset of
chaos\footnote{Small perturbations to the entire system give rise to
structures in the phase space, first discussed by Poincare
\cite{poincare1892} and usually termed ``homoclinic tangles'', that
are quantifiable signatures of chaos.} around the homoclinic orbits
when the system is slightly perturbed from the conservative dynamics,
a fact that could be important in the analysis of the transition to
plunge but which we do not discuss further here.

While homoclinic orbits are present even in comparable mass black hole
systems described in a PN expansion, the complexity of the PN
equations of motion makes analytic results about homoclinic orbits
difficult to come by for comparable mass systems \cite{grossman2008}.
Since most of those references compare results against the Kerr
equatorial case, we restrict our attention here to the fiducial case
of homoclinic orbits in the equatorial plane of Kerr black holes.

\section{Orbital Parameters of Homoclinic Orbits}
\label{physspace}

The equatorial homoclinic Kerr orbits asymptotically approach the
same unstable circular orbit in the infinite future and past, whirling
an infinite number of times as they do so.  In this section, we
provide the afore-promised formal definition of a homoclinic orbit and
substantiate this claim.

\subsection{Definition of a Homoclinic Orbit}
\label{sec:defn}

Formally, a homoclinic orbit approaches the same invariant set in the
infinite future as in the infinite past.  A collection of points $S$
in the phase space of a dynamical system is an invariant set if orbits
that are in the set at any time remain in the set for all previous and
subsequent times.  Of course, the set of points in phase space traced
out by any solution to the equations of motion constitutes an
invariant set, but useful information about global properties of the
phase space usually comes from identifying invariant sets with some
associated \emph{recurrence property}, such as fixed points, periodic
orbits, or the $n$-dimensional tori on which bounded quasiperiodic
motion in integrable systems unfolds.  Henceforth, when we refer to an
invariant set, we will always mean a recurrent invariant set.

The set of all trajectories that approach $S$ asymptotically in the
infinite future is a submanifold of the phase space, namely the stable
manifold of $S$.  Likewise, all trajectories that approach $S$
asymptotically in the infinite past form the unstable manifold of $S$.
A invariant set is called hyperbolic if it has both a stable and an
unstable manifold.

Now, stable and unstable manifolds of invariant sets can sometimes
intersect: some individual trajectories may approach (possibly
different) invariant sets both as $t \to +\infty$ and as $t \to
-\infty$.  When such a trajectory lies in the stable manifold of one
invariant set $S_{\sss{+}}$ and the unstable manifold of a different
invariant set $S_{\sss{-}}$, the trajectory is heteroclinic to
$S_{\sss{+}}$ and $S_{\sss{-}}$.  If instead the trajectory approaches
the same invariant set $S$ in the infinite future and past, i.e. if it
is an intersection of the stable and unstable manifolds of \emph{the
same set} $S$, then the trajectory is homoclinic to $S$.

Identifying the homoclinic orbits in a dynamical system thus amounts
to finding the intersections of the stable and unstable manifolds of
its hyperbolic invariant sets. As we will now show, for the system of
Kerr equatorial orbits, the only hyperbolic invariant sets with
associated homoclinic orbits are the energetically bounded, unstable
circular orbits. Strictly speaking, no relativistic orbits are truly
recurrent since time itself is a coordinate in a relativistic phase
space \cite{schmidt2002,Hinderer:2008dm} and all orbits are unbounded
in their forward motion in time.  We will go to some trouble in paper
II \cite{perez-giz2008} to reduce to a 6D phase space of spatial
coordinates and their conjugate momenta in which circular orbits are
truly recurrent invariant sets.

\subsection{Kerr Equations of Motion}
\label{subsec:eom}

The Kerr metric in Boyer-Lindquist coordinates and geometrized units
($G = c =1$) is
\begin{alignat}{1}
\label{eq:metric}
  \begin{split}
    ds^2 &= 
    - \left( 1-\frac{2Mr}{\Sigma} \right) dt^2
    - \frac{4Mar\sin^2\theta}{\Sigma} dt d\varphi \\
    &\mrph{=}
    {}+ \sin^2\theta
    \left(
    r^2+a^2 + \frac{2Ma^2r\sin^2\theta}{\Sigma}
    \right) d\varphi^2 \\
    &\mrph{=}
    {}+ \frac{\Sigma}{\Delta}dr^2
    + \Sigma d\theta^2
  \end{split}
  \quad ,
\end{alignat}
where $M, a$ denote the central black hole mass and spin angular
momentum per unit mass, respectively, and
\begin{alignat}{1}
\label{eq:DeltaSigmadefs}
  \begin{split}
    \Sigma &\equiv r^2+a^2\cos^2\theta \\
    \Delta &\equiv r^2-2Mr+a^2
  \end{split}
  \quad .
\end{alignat}

Motion along geodesics of (\ref{eq:metric}) conserves orbital energy
$E$, axial angular momentum $L_z$, the Carter constant $Q$
\cite{carter1968}, and of course the rest mass $\mu$ of the test
particle itself.\footnote{$E$ and $L_z$ are associated with
$t$-translation and $\varphi$-translation Killing vectors of the Kerr
metric.  $Q$ is associated with a Killing tensor with a less obvious
geometric interpretation.  In the weak field, $Q \approx L_x^2 +
L_y^2$.}  Since there are as many constants of motion as degrees of
freedom, the usually second order geodesic equations can be integrated
to yield a set of 4 first order equations of motion for the
coordinates \cite{carter1968}.  Before writing them down, we adopt the
useful and now common convention \cite{schmidt2002} to set both $M$
and $\mu$ equal to 1, an operation tantamount to working in units in
which the coordinates $r$ and $t$, the proper time $\tau$, the spin
parameter $a$ and the conserved quantities $E, L_z, Q$ are all
dimensionless.  In these dimensionless units, which we use in the
remainder of this paper, the first-order geodesic equations are
\begin{subequations}
\label{eq:dimcarter}
\begin{alignat}{1}
\Sigma \dot r &= \pm \sqrt{R}
\label{subeq:dimcarter-r}\\
\Sigma \dot \theta &= \pm \sqrt{\Theta}
\label{subeq:dimcarter-theta}\\
\Sigma \dot \varphi &=
\frac{a}{\Delta} \lf( 2rE - aL_z \rt)
+ \frac{L_z}{\sin^2 \theta}
\qquad \quad \elpunc{,}
\label{subeq:dimcarter-phi}\\
\Sigma \dot t&=
\frac{(r^2 + a^2)^2 E - 2arL_z}{\Delta}
- a^2 E \sin^2 \theta
\label{subeq:dimcarter-t}
\end{alignat}
\end{subequations}
where an overdot denotes differentiation with respect to the
particle's (dimensionless) proper time $\tau$ and
\begin{alignat}{1}
  \Theta(\theta) &= Q - \cos^2\theta
  \left\{
  a^2(1- E^2) + \frac{L_z^2}{\sin^2\theta}
  \right\}
  \label{eq:Thetaeq}\\
  \begin{split}
    R(r) &= -(1 - E^2)r^4 + 2r^3 - \lf[ a^2(1 - E^2) + L_z^2 \rt]r^2 \\
    &\mrph{=}  {}+ 2(aE - L_z)^2\, r - Q \Delta
    \qquad \qquad .
  \end{split}
  \label{eq:Rpoly}
\end{alignat}

Modulo initial conditions, we can identify any orbit around a black
hole of some mass and spin by its constants of motion $E,L_z$ and $Q$.
From this point on, we will restrict attention to equatorial
orbits. Equatorial Kerr orbits have $\theta = \pi/2$ and $\dot{\theta}
= 0$.  It follows from the equations of motion (\ref{eq:dimcarter})
that equatorial orbits always have $Q=0$ and that they remain
equatorial and form a self-contained set.

In the following section we will identify the homoclinic orbits with
the aid of an effective potential picture.

\bigskip
\subsection{Effective Potential and the Homoclinic Orbits}

To clarify terms, it is standard parlance to refer to ``unstable''
circular orbits in the black hole system. Strictly speaking, the
unstable circular orbits are actually hyperbolic -- they possess both
a stable and unstable manifold.  Nonetheless, we continue with this
conventional parlance to avoid unnecessarily elaborate verbiage and 
assume the reader understands
phrasing such as ``the stable manifold of an unstable circular
orbit''.

The hyperbolic invariant sets in the equatorial Kerr system are
precisely these unstable circular orbits. Of those, the ones that are
energetically bound ($E < 1$) give rise to homoclinic orbits.

Identification of the homoclinic orbits, and indeed interpretation of
the dynamics in general, is easiest with an effective potential
formulation, which motion around spinning black holes admits.
However, as we explain shortly, the Kerr effective potential has some
awkward features that make our exposition a bit cumbersome.  Thus, to
ease discussion, we briefly recount the effective potential picture
for Schwarzschild black holes
\cite{wald1984,bombelli1992,hartle,chandrasekhar1983} and identify the
homoclinic orbits for that case before extending to Kerr black holes.
This subsection amounts to a synopsis of the familiar specifics of
orbits admitted by the Schwarzschild and Kerr effective potentials, a
lengthier accounting of which we include in Appendix \ref{app:eff} for
the detail conscious reader.

For Schwarzschild orbits ($a=0$), the suggestive form
\begin{equation}
\label{eq:rdotsuggestive}
  \half \dot r^2-\frac{R(r)}{2\Sigma^2} =0
\end{equation}
of the radial equation (\ref{subeq:dimcarter-r}) becomes the familiar
\cite{wald1984,hartle}
\begin{equation}
\label{eq:Veffeom}
  \half \dot r^2+V_{\text{eff}}=\varepsilon_{\text{eff}}
\end{equation}
describing motion in the one-dimensional effective potential
\begin{equation}
\label{eq:VeffSch}
  \begin{split}
    V_{\text{eff}}(r,a=0) &\equiv
    -\frac{R(r)}{2r^4}\evalat{a=0} + \half E^2 \\
    &= -\frac{1}{r} + \frac{L_z^2}{2r^2} - \frac{L_z^2}{r^3} + \half
  \end{split}
  \quad ,
\end{equation}
with effective energy $\varepsilon_{\text{eff}} = \half E^2$.  Note
that the asymptotic value of the potential at $r = \infty$ is $1/2$,
so that $E = 1$ divides bound from unbound motion.

\begin{figure}[hbt]
  \centering
  \hspace{-10mm}
  \begin{minipage}{94mm}
    \includegraphics[width=1.1\textwidth]{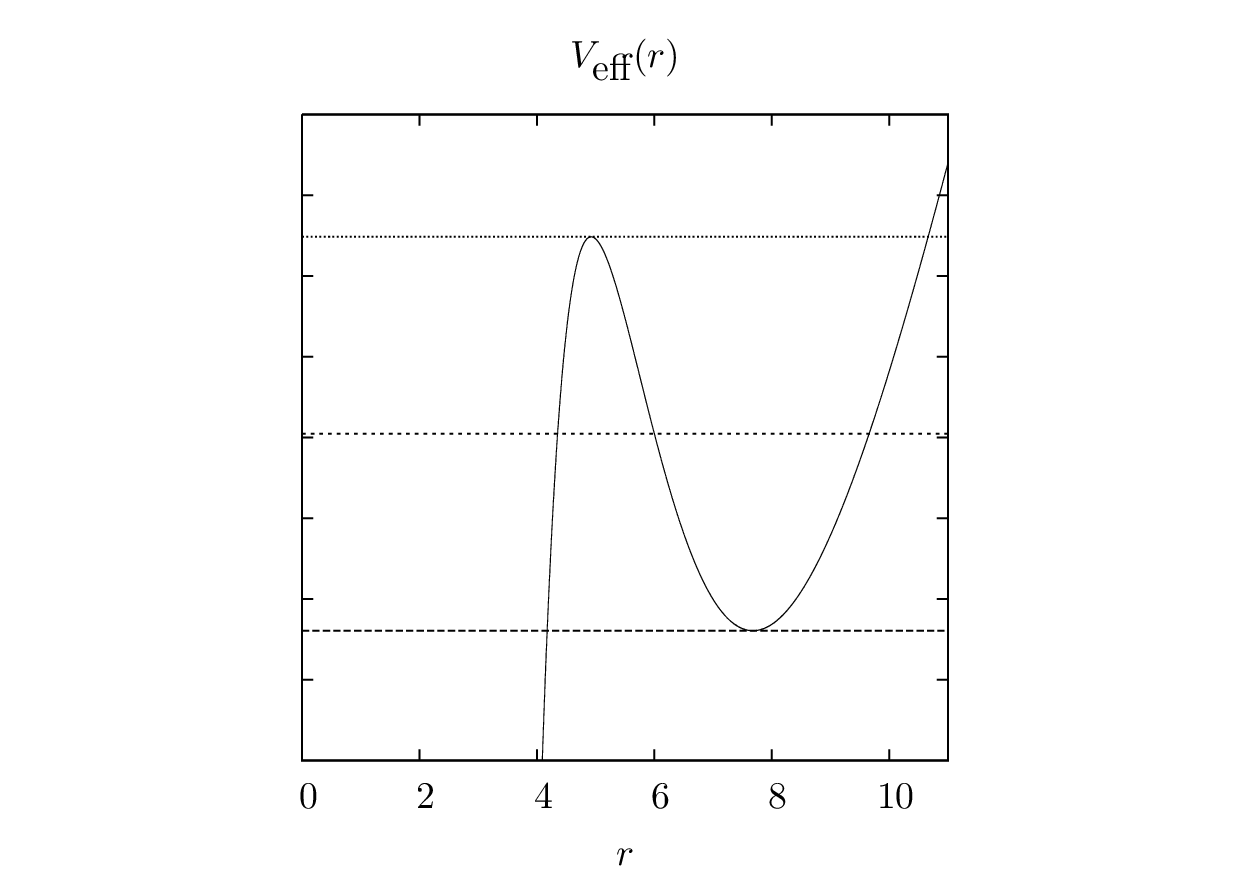}
    \\
    \includegraphics[width=1.1\textwidth]{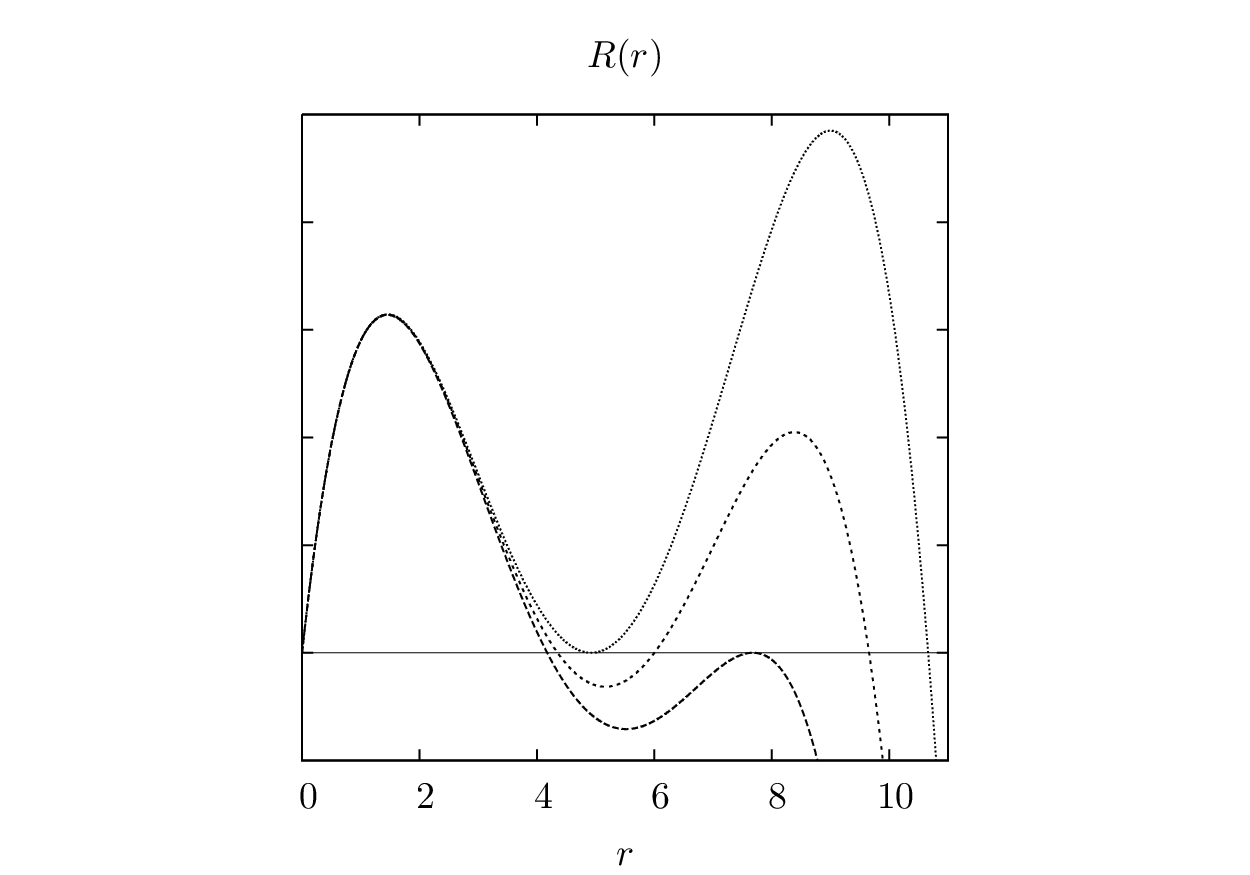}
    \hfill
  \end{minipage}
  \hfill
  \caption{$R(r)$ functions and $V_{\textrm{eff}}(r)$
  (Eq. (\ref{eq:VeffSch})) for 3 Schwarzschild orbits with $L_z =
  3.55$.  From bottom to top in both diagrams, the corresponding
  energies are $E = 0.947421, 0.948707$ and $0.949993$.  The first
  value is for a stable circular orbit at $r_s = 7.679020$, the second
  for an eccentric orbit with $r_p = 6.000593$ and $r_a = 9.656613$,
  and the third is for both an unstable circular orbit at $r_u =
  4.923479$ and for a homoclinic orbit with $r_p = r_u$ and $r_a =
  10.662889$. The vertical scales have been suppressed for visual
  clarity.  Note that $R(r) = R'(r) = 0$ at the circular orbits.}
  \label{RvsVSchFig}
\end{figure}

An example of such an effective potential for a nonspinning black hole
with $L_z=3.55$ is shown in Fig.\ \ref{RvsVSchFig}. It is simple to
read from this figure that the maximum of the potential
($dV/dr=0$,$d^2V/dr^2<0$) corresponds to an unstable circular orbit
and the minimum of the potential ($dV/dr=0$,$d^2V/dr^2>0$) corresponds
to a stable circular orbit.  Note that the energy of the maximum $E_u$ is below
the asymptotic value $E=1$.

As indicated by the solid line, there is another orbit with energy
$E_u$ but an apastron given by the outer intersection of the
horizontal line of energy $E_u$ with $V_{\text{eff}}$.  When released
from rest at the apastron $r_a$, a test particle will roll toward the
unstable circular orbit taking an infinite amount of time to reach the
peak, and likewise if time reversed.  This orbit is a homoclinic
orbit.  For every bound unstable circular orbit there exists such a
homoclinic orbit with the same $E$ and $L_z$.\footnote{There are no
bound orbits with the same $(E,L_z)$ of {\it unbound}, unstable
circular orbits (i.e. those with $E>1$) and therefore the unbound
circular orbits do not possess homoclinic orbits, as elaborated in
appendix \ref{app:eff}.}  Appendix \ref{app:eff} shows that these are
the only homoclinic orbits.

For Kerr black holes ($a\ne0$), the $E$ and $L_z$ dependences in
equation (\ref{eq:rdotsuggestive}) do not separate as they do in the
Schwarzschild case.  The radial motion can still be cast in the form
(\ref{eq:Veffeom}) as the one-dimensional motion of a particle with energy
\begin{subequations}
\label{eq:VeffEffKerr}
\begin{alignat}{1}
  \varepsilon_{\text{eff}} &= 0\\
  \intertext{moving in a potential}  
  V_{\text{eff}}(r) &\equiv -R(r)/2\Sigma^2
  \quad ,
  \label{eq:VeffKerr}
\end{alignat}
\end{subequations}
but unlike when $a =0$, the potential depends on both $E$ and $L_z$
through $R(r)$.  $V_{\text{eff}}$ is thus a different potential for
each orbit (i.e. for each $(E, L_z)$ pair) instead of a single
potential for an entire family of orbits like $V_{\text{eff}}(a=0)$.
Fig.\ \ref{RvsVSchFig} plots various such functions $R(r)$ in the
lower panel. As the figure highlights, having the potential vary under
one's feet, so to speak, as the energy of the particle changes means
that information we could previously glean from a single plot of
$V_{\text{eff}}(a=0)$ is now diluted over an infinite number of plots
of $R(r)$.  Nevertheless, a bit more effort -- expended in appendix
\ref{app:eff} -- shows that even when the black hole spins the unstable circular
orbits are still the only hyperbolic invariant sets and that those
with $E < 1$ give rise to homoclinic orbits.

Although $R(r)$ changes with $E$, we can still read qualitative
features of the motion effectively from a plot of $R(r)$.  To clarify
the visual interpretation, Fig.\ \ref{RvsVSchFig} plots the $R(r)$ for
$a =0$ and $L_z=3.55$ below a plot of the corresponding Schwarzschild
$V_{\textrm{eff}}(r)$.  Whereas in the effective potential diagram the
$r$ values accessible to a particle with given energy are those for
which $V_{\textrm{eff}}$ is below the constant energy line, in the
pseudo-potential diagram of a given orbit the accessible $r$ values
are those for which $R(r)$ is above the zero line, reflecting the fact
that $\dot r$ in Eq.\ (\ref{eq:dimcarter}) is real so that the $R(r)$
under the radicand must be non-negative.

Turning points of the motion, for which $V_{\text{eff}}(r) =
\varepsilon_{\text{eff}} = 0$, correspond to single roots of $R(r)$.
Circular orbits require both $V_{\text{eff}} = 0$ and
$\txtD{r}{V_{\text{eff}}} = 0$, or the equivalent
\begin{alignat}{3}
\label{eq:circcond}
  R(r) &= 0 & \quad &\text{and} \quad & R'(r) &= 0
  \quad ,
\end{alignat}
and thus correspond to double roots of $R$, as Fig.\ \ref{RvsVSchFig}
confirms.  Simultaneously solving these equations yields expressions,
originally published in Ref.\
\cite{Bardeen1972},
\begin{subequations}
\label{eq:ELcirc}
\begin{align}
    E &= \ph{\pm} \frac{r^{3/2} - 2r^{1/2} \pm a}
    {r^{3/4}\sqrt{r^{3/2} - 3r^{1/2} \pm 2a}}
    \label{eq:Ecirc}\\
    L_z &= \pm \frac{r^2 \mp 2ar^{1/2} + a^2}
    {r^{3/4}\sqrt{r^{3/2} - 3r^{1/2} \pm 2a}}
    \label{eq:Lcirc}
\end{align}
\end{subequations}
for the energy and angular momentum of circular orbits.
The top/bottom signs denote prograde/retrograde.

Two noteworthy circular orbits deserve mention: the innermost stable
circular orbit (isco) and the innermost bound circular orbit
(ibco). As the angular momentum decreases, the stable and unstable
circular orbits merge to a saddle point -- the isco.  It is the
circular orbit for which $E$ and $\absval{L_z}$ are a
minimum\footnote{
Statements that apply to both prograde and retrograde trajectories are
phrased in terms of $\absval{L_z}$.}
\cite{Bardeen1972}:
\begin{alignat}{1}
\label{eq:Kerrisco}
  r_{\text{isco}} &= 3 + Z_2 \mp \sqrt{(3 - Z_1)(3 + Z_1 +2Z_2)} \\
  Z_1 &\equiv 1 + \sqrt[3]{1 - a^2}
  \left[ \sqrt[3]{1 + a} + \sqrt[3]{1 - a} \right] \nonumber \\
  Z_2 &\equiv  \sqrt{3a^2 + Z_1^2} \nonumber
  \quad .
\end{alignat}
Since $R''(r_{\text{isco}}) = 0$ when $E=E_{\text{isco}},
\absval{L_z}=\absval{L_{\text{isco}}}$, the isco corresponds to the
only possible triple root of $R$.  The ibco is the marginally bound
$E=1$, unstable circular orbit \cite{Bardeen1972}:
\begin{equation}
\label{eq:ribcodef}
  r_{\text{ibco}} \equiv 2 \mp a + 2\sqrt{1 \mp a}
  \quad .
\end{equation}

The upshot is that every $\absval{L_{\text{isco}}} < \absval{L_z} <
\absval{L_{\text{ibco}}}$ admits a bound unstable circular orbit and a
corresponding homoclinic orbit with the same $(E, L_z)$.  The apastron
of the homoclinic orbit with $(E_{\text{ibco}},L_{\text{ibco}})$ is
$r_a=\infty$ while the apastron (and periastron) of the homoclinic
orbit with $(E_{\text{ibco}},L_{\text{ibco}})$ is
$r_a=r_{\text{isco}}$. In other words, the ibco has a homoclinic orbit
with eccentricity 1 and the isco \emph{is} a homoclinic orbit with
eccentricity zero.  The eccentricities of the homoclinic orbits range
from $1$ down to $0$.

\subsection{Exact expressions for orbital parameters of homoclinic orbits}

Above we have described the homoclinic orbits by their $E$ or
$L_z$.
There are other ways to describe the homoclinic orbits.
In general, 
non-circular equatorial Kerr orbits form a two-parameter set,
with any particular orbit specified by its
energy and angular momentum.  For bound non-plunging orbits, other
pairs of independent orbital parameters can also be used, such as the
periastron and apastron $(r_p,r_a)$ or, as is often done,
appropriately defined pseudo-Keplerian parameters $(e,p)$
(eccentricity and semi-latus rectum, respectively).  Homoclinic
orbits, however, lie in one-to-one correspondence with the $E_u < 1$
unstable circular orbits, a one-parameter family specified by the
radius $r_u$.  Homoclinic orbits thus form a one-parameter
family all of whose orbital parameters depend only on the single
parameter $r_u$.

For $E$ and $L_z$ this is clearly the case -- homoclinic orbits have
the same energy and angular momentum as the circular orbit they
asymptotically approach, and equations (\ref{eq:ELcirc}) determine $E$
and $L_z$ once $r_u$ is specified.  Homoclinic orbits also form the
separatrix between plunging and non-plunging orbits, so they, like any
bound non-plunging orbit, have well-defined values of $r_p, r_a, e,
p$.  Simple expressions for those parameters follow from rewriting the
$R(r)$ function, which has a double root at $r_u$ for homoclinic
orbits, as
\begin{equation}
\label{eq:Reqrura}
  R(r) = (E^2 - 1)\, r (r - r_u)^2 (r - r_a)
  \quad ,
\end{equation}
where $r_a$ is the apastron of the homoclinic orbit.  Expanding
(\ref{eq:Reqrura}) and equating powers of $r$ with equation
(\ref{eq:Rpoly}) for $R(r)$ yields relations among $r_u, r_a, E$ and
$L_z$.  In particular, equality of the linear coefficients implies
that
\begin{equation}
  r_a= \frac{2 (aE - L_z)^2}{(1 - E^2) r_u^2}
  \quad .
\end{equation}
Substituting $E(r_u)$ and $L_z(r_u)$ from (\ref{eq:ELcirc}) for $E$
and $L_z$ and simplifying leads to the expression
\begin{equation}
\label{eq:rahomo}
  r_a = \frac{2 r_u (a \mp
  \sqrt{r_u})^2}{r_u^2 - 4r_u \pm 4a\sqrt{r_u} - a^2}
\end{equation}
for the apastron of a homoclinic orbit.

Eq.\ (\ref{eq:rahomo}) also furnishes expressions for the $e$ and $p$ of
a homoclinic orbit in terms of $r_u$.  In analogy with Keplerian
orbits, the eccentricity\footnote{Note from (\ref{eq:eccdef}) that $e$
varies from 0 (for circular orbits, whose $r_a = r_p$) to 1 (for
orbits with $E \to 1$, whose $r_a \to \infty$).} and semi-latus
rectum of a generic orbit are typically defined via
\begin{alignat}{2}
\label{eq:eccpimpdefs}
  r_p &\equiv \frac{p}{1 + e}\,, \quad & r_a &\equiv \frac{p}{1 - e}
  \quad ,
\end{alignat}
or equivalently
\begin{alignat}{1}
  e &\equiv \frac{r_a - r_p}{r_a + r_p}
  \label{eq:eccdef}\\
  p &\equiv \frac{2 r_a r_p}{r_a + r_p}
  \label{eq:pdef}
  \quad .
\end{alignat}
Substituting (\ref{eq:rahomo}) into (\ref{eq:eccdef}) and
(\ref{eq:pdef}) with $r_p = r_u$ yields
\begin{alignat}{1}
  e^{\text{hc}} &=
  \frac{-r_u^2 + 6r_u \mp 8a\sqrt{r_u} + 3a^2}{r_u^2 - 2r_u + a^2}
  \label{eq:ecchomo}\\
  p^{\text{hc}} &=
  \frac{4 r_u \lf(a \mp \sqrt{r_u} \rt)^2}{r_u^2 - 2r_u + a^2}
  \qquad\qquad \elpunc{.}
  \label{eq:phomo}
\end{alignat}
Ref.\ \cite{O'Shaughnessy:2002ez} derives the implicit relation
\begin{equation}
\label{eq:pehomoimplicit}
  \begin{split}
    0 &= p^2 (p - 6 - 2e)^2 + a^4 (e - 3)^2 (1 + e)^2 \\
    &\mrph{=} {}- 2a^2 p (1 + e) \lf[ 14 + 2e^2  + p(3 - e) \rt]
  \end{split}
\end{equation}
that $e$ and $p$ of the homoclinic orbit (referred to there as ``the
separatix'') must satisfy, and an equivalent implicit expression also
appears in \cite{glampedakis2002, Glampedakis:2005hs}.  Acknowledging
the relationship between the homoclinic orbits and unstable circular
orbits from the outset furnishes the explicit parametric solutions
(\ref{eq:ecchomo}) and (\ref{eq:phomo}) to those implicit equations.

We now know how to specify the equatorial circular orbits by a single
parameter; either $E$ or $L_z$ for instance. The unstable circular
orbits are a family of hyperolic sets, which means they have stable
and unstable manifolds. We have also derived the perihelia and apastra of the
homoclinic orbits as well as the $(e,p)$ as explicit functions of
$r_u$ and spin. 

However, we can
do better than this. We can find exact solutions for the
homoclinic trajectories themselves as a function of spin. We will do this 
now.

\section{Exact solutions for equatorial homoclinic orbits}
\label{exact}

 
An exact solution to geodesic motion is a rare commodity. In this
section we very briefly sketch the derivation of an exact parameteric
solution for homoclinic orbits around Kerr black holes of arbitrary
spin and refer the reader to the acrobatics of appendix \S
\ref{append1} for the detailed derivation.

For any equatorial orbit, the radial motion consists of alternating
inbound phases ($\txtD{\tau}{r} < 0$) and outbound ($\txtD{\tau}{r} >
0$) phases during which $r$ varies monotonically with time.  Because
the equations of motion depend explicitly only on $r$, the radial
coordinate parametrizes the motion during any single such phase, and
other dynamical variables can be expressed in terms of $r$.

Consequently, during an inbound phase, the integrated proper time,
coordinate time, and azimuth between some reference point $r_0$ and
$r$ are
\begin{widetext}
\begin{alignat}{2}
  \tau(r) &= -\int^r_{r_0} dr \D{r}{\tau} &&= 
  -\int^r_{r_0} dr \frac{\Sigma}{\sqrt{R}}
  \label{eq:taureq}\\
    t(r) &= -\int^r_{r_0} dr\,
    \frac{\txtD{\tau}{t}}{\txtD{\tau}{r}}
    &&= -\int^r_{r_0} dr\,
    \frac{r^2(r^2 + a^2)E + 2a(aE - L_z)r}{\Delta\sqrt{R}} \quad ,
  \label{eq:treq}\\
    \varphi(r) &= -\int^r_{r_0} dr\,
    \frac{\txtD{\tau}{\varphi}}{\txtD{\tau}{r}}
    &&= -\int^r_{r_0} dr\,
    \frac{r^2L_z + 2(aE - L_z)r}{\Delta\sqrt{R}}
  \label{eq:phireq}
\end{alignat}
\end{widetext}
where $r$ and $r_0$ are both radial coordinates along the same phase
(i.e. along a given half-leaf) of the motion.  Removing the overall
minus signs yields the corresponding expressions for outbound motion.
Eqs.\ (\ref{eq:taureq})-(\ref{eq:phireq}) and their outbound
counterparts are correct for both $r < r_0$ and $r > r_0$ along a
single inbound/outbound phase.

For ordinary eccentric orbits, $R(r)$ has four distinct roots and
equations (\ref{eq:taureq}) - (\ref{eq:phireq}) are at best elliptic
integrals.  However, the fact that $R$ factors as in
(\ref{eq:Reqrura}) for homoclinic orbits renders the integrals soluble
in terms of elementary functions.  We integrate these equations
analytically in appendix \ref{append1} to give:
\begin{widetext}
\begin{subequations}
\label{eq:exact}
\begin{align}
  \begin{split}
    \tau(r) &= \frac{1}{\sqrt{1 - E^2}}\sqrt{r(r_a - r)}
    + \frac{2}{\lf(1 - E^2\rt)^{3/2}}
    \atan\sqrt{\frac{r_a - r}{r}}
  + \frac{2}{\gamma\lambda_r}
    \atanh\sqrt{\frac{r_u}{r_a - r_u}\frac{r_a - r}{r}}    
  \end{split}
\label{eq:exact-tau}
\\
  \begin{split}
    t(r) &= \frac{E}{\sqrt{1 - E^2}}\sqrt{r(r_a - r)}
  + 2E\frac{\lf(3 - 2E^2\rt)}{\lf(1 - E^2\rt)^{3/2}}
    \atan\sqrt{\frac{r_a - r}{r}}
 + \frac{2}{\lambda_r}
    \atanh\sqrt{\frac{r_u}{r_a - r_u}\frac{r_a - r}{r}}
     \\
    &\quad \quad - \frac{2 r_{\sss +}}{\sqrt{1 - a^2}}
    \atanh
    \sqrt{
      \frac{r_{\sss +}}{r_a - r_{\sss +}}
      \frac{r_a - r}{r}
    } 
- \frac{2 r_{\sss -}}{\sqrt{1 - a^2}}
    \atanh
    \sqrt{
      \frac{r_{\sss -}}{r_a - r_{\sss -}}
      \frac{r_a - r}{r}
    }
  \end{split}
\label{eq:exact-t}
\\
  \begin{split}
    \varphi(r) &= 2 \frac{\Omega_u}{\lambda_r}
    \atanh\sqrt{\frac{r_u}{r_a - r_u}\frac{r_a - r}{r}}
- \frac{a}{\sqrt{1 - a^2}}
    \atanh
    \sqrt{
      \frac{r_{\sss +}}{r_a - r_{\sss +}}
      \frac{r_a - r}{r}
    } 
    - \frac{a}{\sqrt{1 - a^2}}
    \atanh
    \sqrt{
      \frac{r_{\sss -}}{r_a - r_{\sss -}}
      \frac{r_a - r}{r}
    }
  \end{split}
  \quad ,
\label{eq:exact-phi}
\end{align}
\end{subequations}
\end{widetext}
where we have set $\tau = t = \varphi = 0$ at $r = r_a$, the apastron
(\ref{eq:rahomo}) of the homoclinic orbit.

In the equations above, $r_{\sss{+}}$ and $r_{\sss{-}}$ represent,
respectively, the outer and inner horizons of the black hole, $\gamma
\equiv \txtD{\tau}{t}(r_u)$ and $\Omega_u \equiv
\frac{d\varphi}{dt}(r_u)$ are the (constant) Lorentz factor and
azimuthal velocity ($\Omega_u > 0$ for prograde orbits, $\Omega_u < 0$
for retrograde) of the associated unstable circular orbit, and $E$ is
the energy of the homoclinic orbit (and also of the unstable circular
orbit).  The remaining parameter is 
\begin{equation}
\lambda_r =
\left. \frac{1}{\gamma\Sigma} \sqrt{\frac{R''}{2}}\right |_{r_u}\quad ,
\end{equation} 
where a prime denotes differentiation with respect to $r$.
As derived in paper II in this series \cite{perez-giz2008},
$\lambda_r$ is the radial stability
exponent of the unstable circular orbit.

Eq (\ref{eq:exact}) reveals an interesting and unobvious fact about
the homoclinic orbits.  Consider a circular orbit at $r=r_u$ with
energy $E$ and $L_z$ and a homoclinic orbit with the same energy and
angular momentum.  Even though the homoclinic orbit takes an infinite
amount of time to asymptote to or away from $r_u$, the total
accumulated phase difference between the homoclinic orbit and the
circular orbit over that infinite period is finite.

\begin{figure*}[htb]
  \centering
  \begin{minipage}{1.1\textwidth}
    \hspace{-.1\textwidth}
    \includegraphics[width=.49\textwidth]{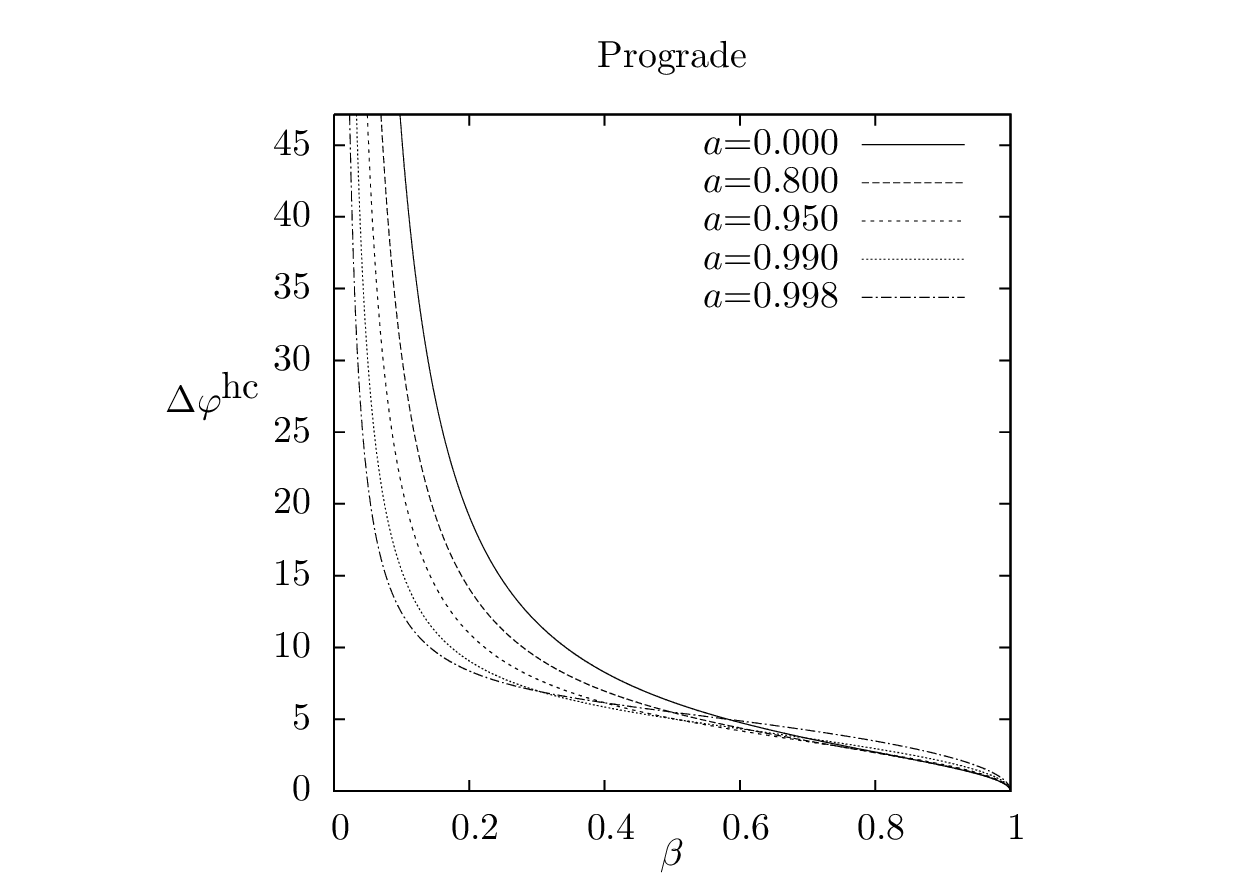}
    \hspace{-.1\textwidth}
    \includegraphics[width=.49\textwidth]{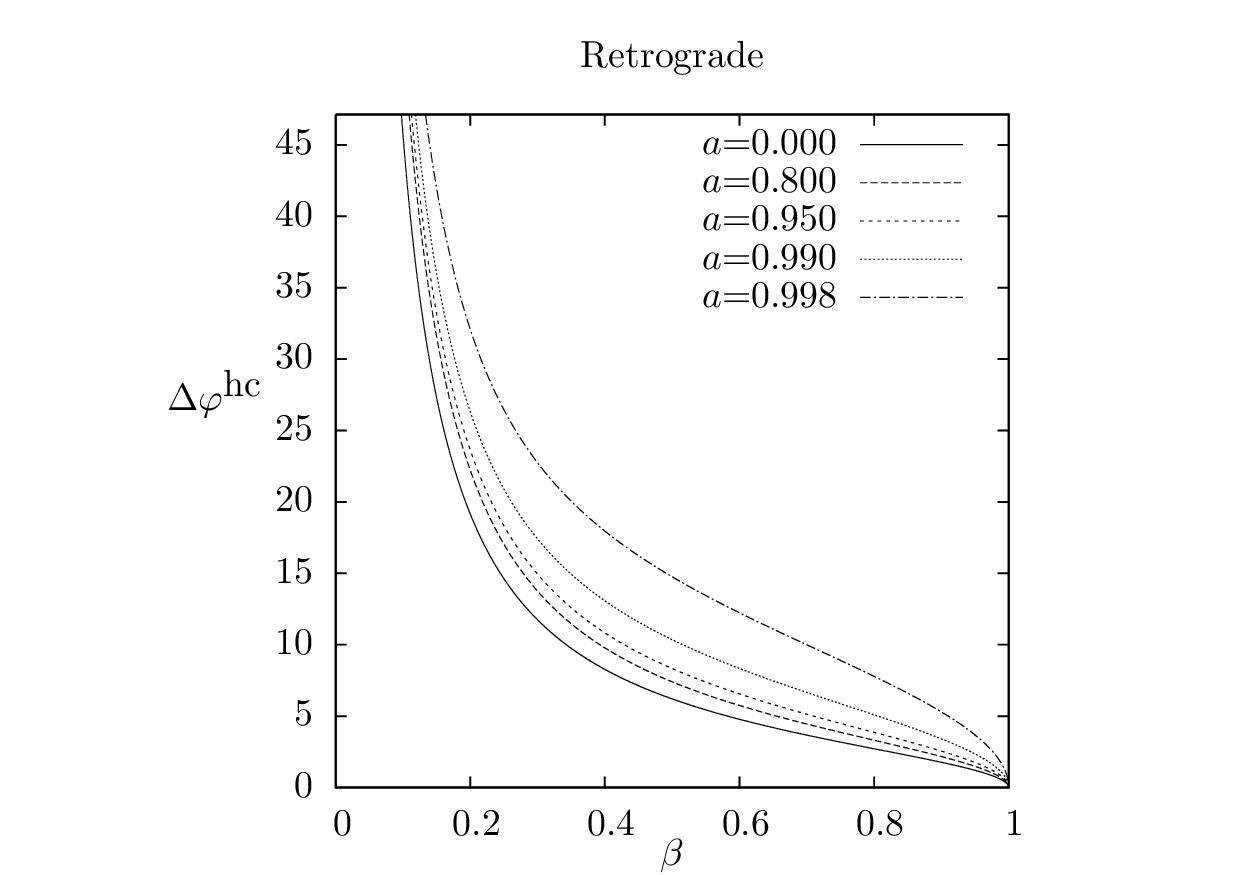}
    \hfill
  \end{minipage}
  \caption{The accumulated phase difference between a homoclinic orbit
  and the unstable circular orbit to which it is doubly asymptotic as
  a function of $r_u$.  For a given spin $a$, the parameter $\beta$
  varies linearly from $r_{\text{ibco}}$ when $\beta = 0$ to
  $r_{\text{isco}}$ when $\beta = 1$. Upper: Prograde homoclinic
  orbits. Lower: Retrograde homoclinic orbits.}
  \label{fig:delphihc}
\end{figure*}

To be concrete, consider a prograde homoclinic orbit with $t=0,
\varphi=0$ at $r=r_a$, and let the circular orbit at $r_u$ be at
$\varphi = 0$ at the same time.  Since $r$ varies monotonically with
$t$ along the homoclinic orbit during its inbound phase (as $t
\approaches \infty$), we can use the $r$ coordinate along the
homoclinic orbit as a global time parameter via Eq.\
(\ref{eq:exact-t}).  Since $\varphi$ along the circular orbit
increases linearly at a rate $\omega_\varphi = \Omega_u$, the phase
difference between the circular and homoclinic orbits is just the
difference between
\begin{equation}
  \varphi^{\text{circ}}(t(r)) = \Omega_u t(r),
\end{equation}
and Eq.\ (\ref{eq:exact-phi}).  By time-reversal symmetry, doubling
this yields the total phase difference between the circular and
homoclinic orbits summed over both the inbound and outbound phases.
Letting $t(r)$ denote time along the inbound ($\dot{r} < 0$) portion of
the homoclinic orbit, the resulting phase difference
\begin{widetext}
\begin{equation}
\label{eq:delphihcdef}
  \begin{split}
    \Delta\varphi^{\text{hc}}(t(r)) &\equiv 
    2 \lf[
      \varphi^{\text{circ}}\lf( t(r) \rt) -
      \varphi^{\text{hc}}\lf( t(r) \rt)
      \rt]\\
    &= 2 \Omega_u \frac{E}{\sqrt{1 - E^2}}
    \lf\{
    \sqrt{r(r_a - r)}+
    2\, \frac{3 - 2 E^2}{1 - E^2} \atan \sqrt{\frac{r_a - r}{r}}
    \rt\}\\
    &\mrph{=} {}+\frac{2}{\sqrt{1 - a^2}}
    \lf\{
    (a - 2\Omega_u r_{\sss{+}})
    \atanh \sqrt{ \frac{r_{\sss{+}}}{r_a - r_{\sss{+}}}
      \frac{r_a - r}{r} }+
    (a - 2\Omega_u r_{\sss{-}})
    \atanh \sqrt{ \frac{r_{\sss{-}}}{r_a - r_{\sss{-}}}
      \frac{r_a - r}{r} }
    \rt\}
  \end{split}
\end{equation}
\end{widetext}
has no divergences, and the limit
\begin{equation}
\label{eq:delphihclim}
  \lim_{t \to \infty} \Delta\varphi^{\text{hc}}(t) =
  \lim_{r \to r_u} \Delta\varphi^{\text{hc}}(r)
\end{equation}
exists.

Fig.\ \ref{fig:delphihc} shows how $\Delta\varphi^{\text{hc}}$ depends
on $r_u$ for various values of the black hole spin $a$.  For ease of
comparison, $\Delta\varphi^{\text{hc}}$ is plotted versus a parameter
that varies linearly from $0$ when $r_u = r_{\text{ibco}}$ to $1$ when
$r_u = r_{\text{isco}}$ for a given $a$.  The fact that
$\Delta\varphi^{\text{hc}} \neq {0\mod 2\pi}$ for all but a measure
zero set of homoclinic orbits means that a generic equatorial
homoclinic orbit asymptotes in the infinite future to a circular orbit
out of phase by $\Delta\varphi^{\text{hc}}$ with the circular orbit
(at the same $r_u$) to which the homoclinic orbit asymptotes in the
infinite past.\footnote{This is why we speak about an orbit being
homoclinic to some invariant set (e.g., the locus of points in phase
space with $r=r_u, p_r = 0, \varphi$ arbitrary) and not about its
being homoclinic to a particular \emph{orbit} (e.g. a particular
unstable circular orbit, including choice of phase).}  Stated another
way, if we were to treat all circular orbits at radius $r_u$ with
different phases as distinct, then by adding a constant and finite
phase to any homoclinic orbit, we could speak meaningfully about
synchronizing it with exactly one such circular orbit at $t=+\infty$
at $r_u$ and with exactly one circular orbit at $t=-\infty$ also at
$r_u$ but with a different phase.

Except for a measure zero set that accumulate a total phase difference
$\Delta\varphi^{\text{hc}} \mod 2\pi = 0$ relative to a circular orbit
over their infinite period motion, a homoclinic orbit orbit that
synchronizes with a given circular orbit at $t = -\infty$ will be out
of phase with that same circular orbit at $t = +\infty$ by
$\Delta\varphi^{\text{hc}}$.  Although a fine detail at this point,
such phase information could be significant to gravational wave
templates for the full black hole spectrum.

We can paint a homoclinic approach to the
unstable circular orbit using the exact solution of this section. We
will do so in the context of the special set of periodic orbits in the
following section.

\section{Homoclinic Limit of Zoom-Whirl Orbits}
\label{whirlim}

Before concluding, we mention another perspective on the physical portrait of the
homoclinic landmark, and that is the connection to zoom-whirl behavior.
An association with zoom-whirl behavior has long been suspected, yet
also subtley misunderstood.
Many practitioners, 
including the present authors, suspected that zoom-whirl behavior was
bound to the proximity to the separatrix.
To the contrary, we found in a previous work \cite{levin2008}, that
zoom-whirl behavior emerges in the strong-field for {\it any}
eccentricity. Put another way, zoom-whirl behavior is demonstrated by
orbits that are {\it not} in the vicinity of the homoclinic orbit as
well as by those that are. Still,
homoclinic orbits do have an important significance as the infinite
whirl limit in the spectrum of zoom-whirl orbits, as we now make explicit.

In Ref.\ \cite{levin2008}, we realized that the spectrum of all black hole
orbits naturally fall into periodic tables -- tables with an infinite
sequence of entries for a given angular momentum around a given black
hole. Each entry in the periodic table is an exactly periodic orbit
characterized by a rational number that immediately identifies the
number of zooms, the number of whirls, and the order in which the
zooms are executed.

\begin{figure*}
  \centering
 \includegraphics[width=0.25\textwidth]{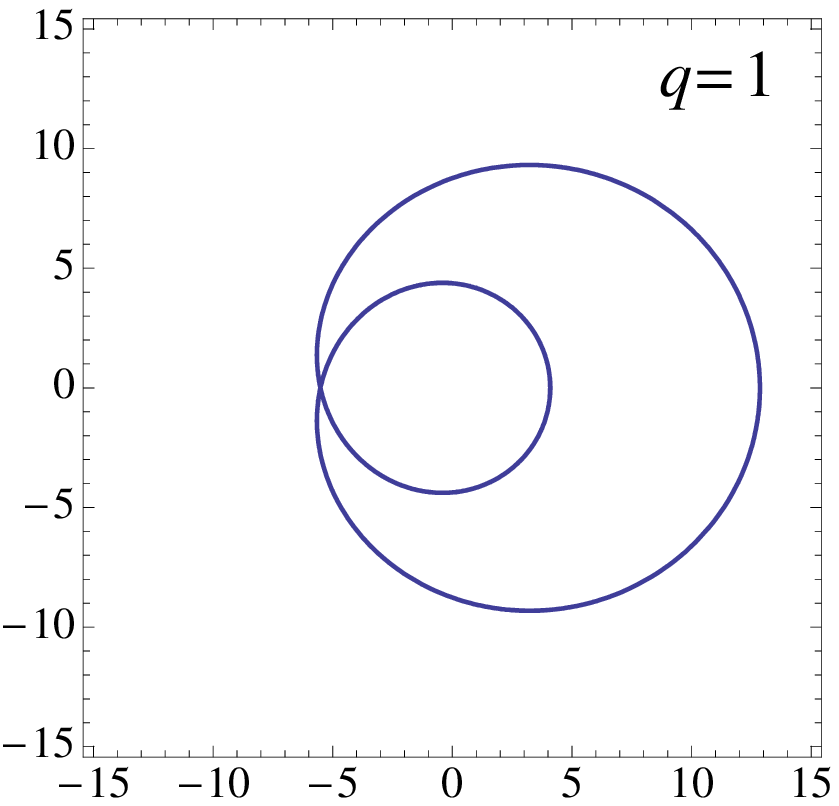}
  \includegraphics[width=0.246\textwidth]{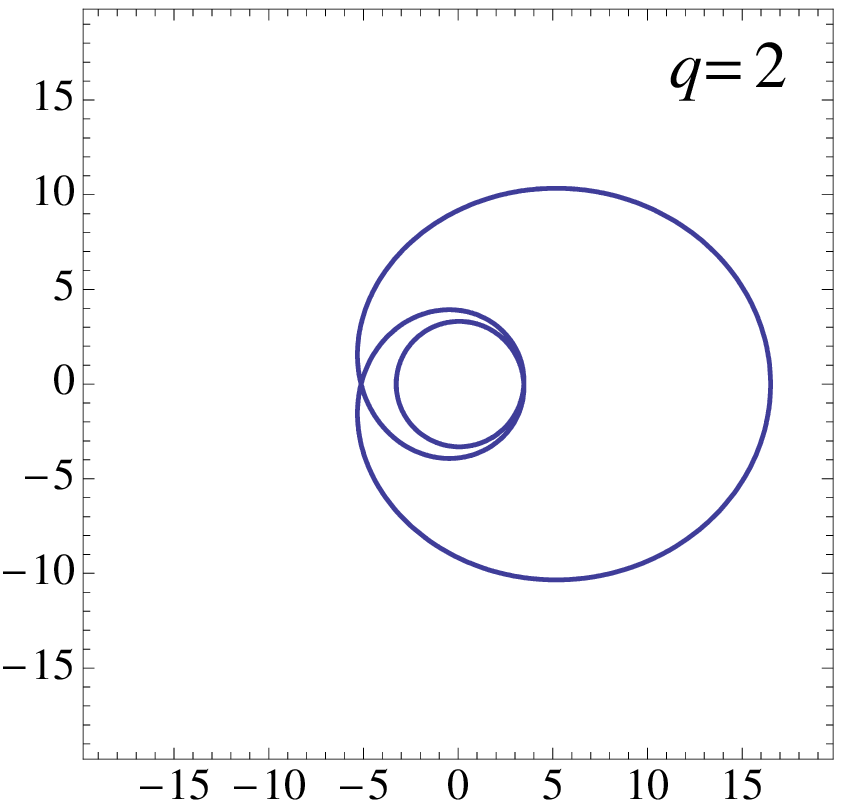}
  \includegraphics[width=0.25\textwidth]{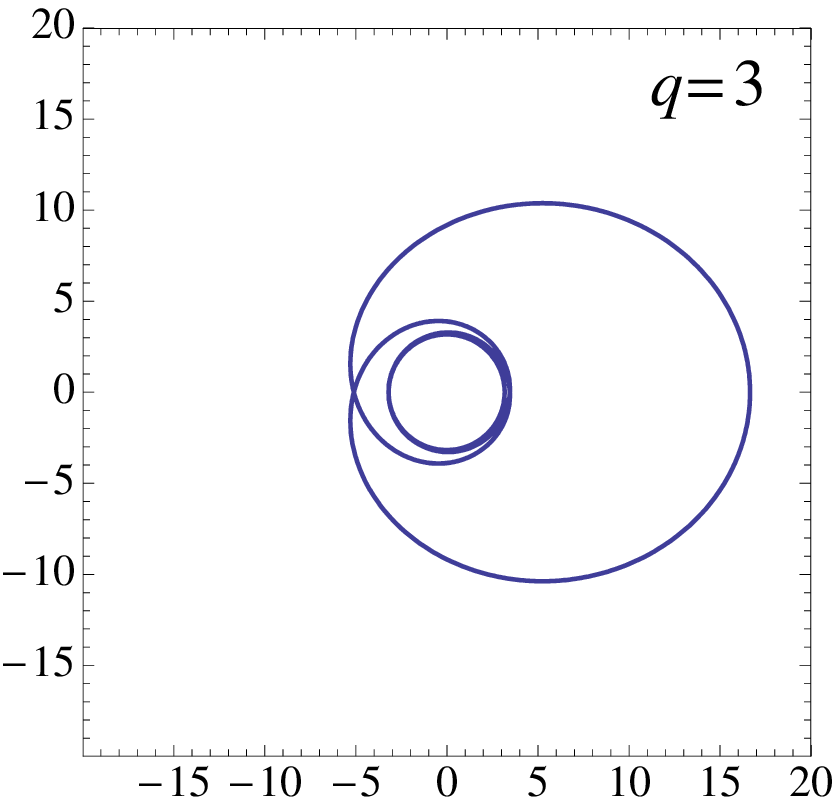}
  \includegraphics[width=0.25\textwidth]{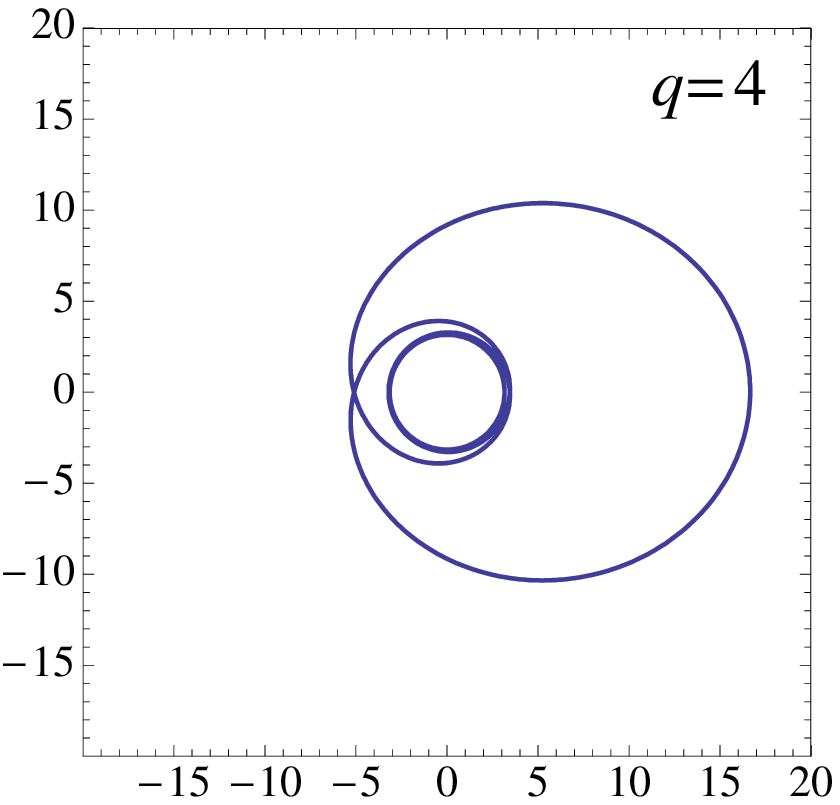}
  \includegraphics[width=0.25\textwidth]{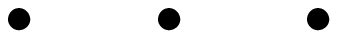}
  \includegraphics[width=0.25\textwidth]{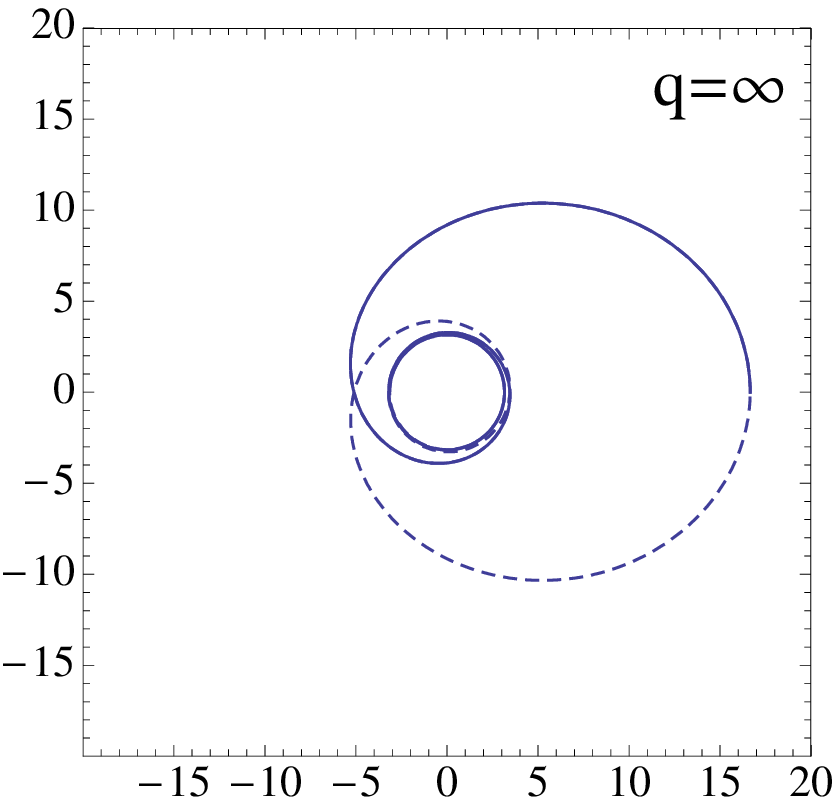}
\hfill
  \caption{The progression of the 1-leaf periodic orbits through $
   1,2,3,4...\infty$ whirls.  The orbits shown are prograde orbits for
   $a=0.5$ and $L_z=3.158540$, or the average of $L_{\text{isco}}$ and
   $L_{\text{ibco}}$.  Note that the whirls beyond the second whirl
   are too closely packed in $r$ to distinguish visually in the plot.}
  \label{fig:z1winflimit}
\end{figure*}

The importance of the periodic tables lies in the observation that
{\it every}
bound orbit can be approximated to arbitrary precision by a periodic
one. Perhaps more important to future studies of gravitational waves, 
{\it every} bound orbit
can be modeled  as a slow precession around
some low-leaf periodic orbit, just as Mercury's orbit can be modeled as a
precession around an ellipse.

Although homoclinic orbits are formally aperiodic in the sense that
they never return to their initial conditions, they nonetheless mimic
periodic orbits, in particular they are the infinite whirl limit in
the periodic sequence -- the final entry in the infinite periodic table
\cite{levin2008}.

To make this connection explicit, consider
single-leaf orbits like those in
Fig.\ \ref{fig:z1winflimit}. Single-leaf orbits are in
one-to-one correspondence with the whole numbers; that is, the
rational associated with each one-leaf orbit counts the integer number
of whirls.  The final entry in this infinite list is the orbit
that executes an infinite number of whirls and therefore never
actually reaches the end of its first radial cycle. That orbit is of
course the homoclinic orbit.

More specifically,
if we write down a sequence of energies $E_w$,
eccentrities $e_w$, apastra ${r_a}_w$, periastra ${r_p}_w$, actions
${J_r}_w$, etc. for the constant $L_z$ set of one-leaf orbits, then
all of these sequences converge in the $w \to \infty$ limit to the
values for the homoclinic orbit with the same $L_z$ (for a black hole
of a given spin).

As already noted, for a given $a$ and $L_z$, the homoclinic orbits
form the separatrix in the phase space between orbits that are
energetically bound and those that are not.  As discussed in detail in
\cite{schmidt2002,levin2008,Hinderer:2008dm}, orbits that are bound in
the phase space turn out to lie on surfaces homeomorphic to
2-dimensional tori (3-tori for generic nonequatorial orbits), and only
these bound orbits have an associated set of fundamental frequencies
in terms of which orbit functionals can be Fourier expanded
\cite{drasco2006}.  The homoclinic orbit of a given $L$ is also
therefore the separatrix between the regions of phase space inhabited
by these quasiperiodic orbits and those that are fully aperiodic.
Since all quasiperiodic orbits can be approximated by the periodic
set, the homoclinic orbit is the divide between the domain of
influence of the periodic set with its correspondence to the rationals
and aperiodic orbits that merge or escape.

\section{Conclusions}
\label{conc}

Homoclinic orbits offer the kind of crucial signpost that demarcates
physically distinct regions of the conservative and inspiral dynamics:
bound from plunging, whirling from not-whirling, smooth from chaotic.
They thereby define salient details of black hole
dynamics and we have spent time in this article deriving
an exact parameteric solution for homoclinic motion that we hope will
prove of use to others in the field.

Physically, we have shown that homoclinic trajectories are an infinite whirl
limit of the zoom-whirl orbits. Even remembering that zoom-whirl
behavior is generic and not exotic in the strong-field
\cite{levin2008}, the homoclinic orbits themselves are a special and
sparse subset. Nonetheless, {\it every} inspiraling orbit must transit
through a homoclinic orbit on the transition to plunge. The isco,
which is the exit to plunge for quasi-circular inspiral, is itself a
homoclinic orbit with eccentricity zero. The homoclinic family ranges
in eccentricity from zero (the isco) all the way up to 1 (homoclinic
to the ibco). {\it All} orbits, except those exceptionally
well-approximated as quasi-circular, will roll through another member
of the homoclinic family on the transition to plunge.

\acknowledgements

We are especially grateful to Becky Grossman for her valuable and
generous contributions to this work.  We also thank Bob Devaney for
helpful input concerning dynamical systems language.  JL and GP-G
acknowledge financial support from a Columbia University ISE
grant. This material is based in part upon work supported under a
National Science Foundation Graduate Research Fellowship.

\vfill\eject

\appendix

\section{The effective potential}
\label{app:eff}
\subsection{$a=0$}

Although the key features of the Schwarzschild geometry are
recognizable at a glance to anyone familiar with an effective
potential formulation, the Kerr case is less visually informative. In
order to ground the details of the circular and homoclinic orbits, we
include the Schwarzschild treatment in detail.

\begin{figure}[h]
  \centering
  \includegraphics[width=0.5\textwidth]{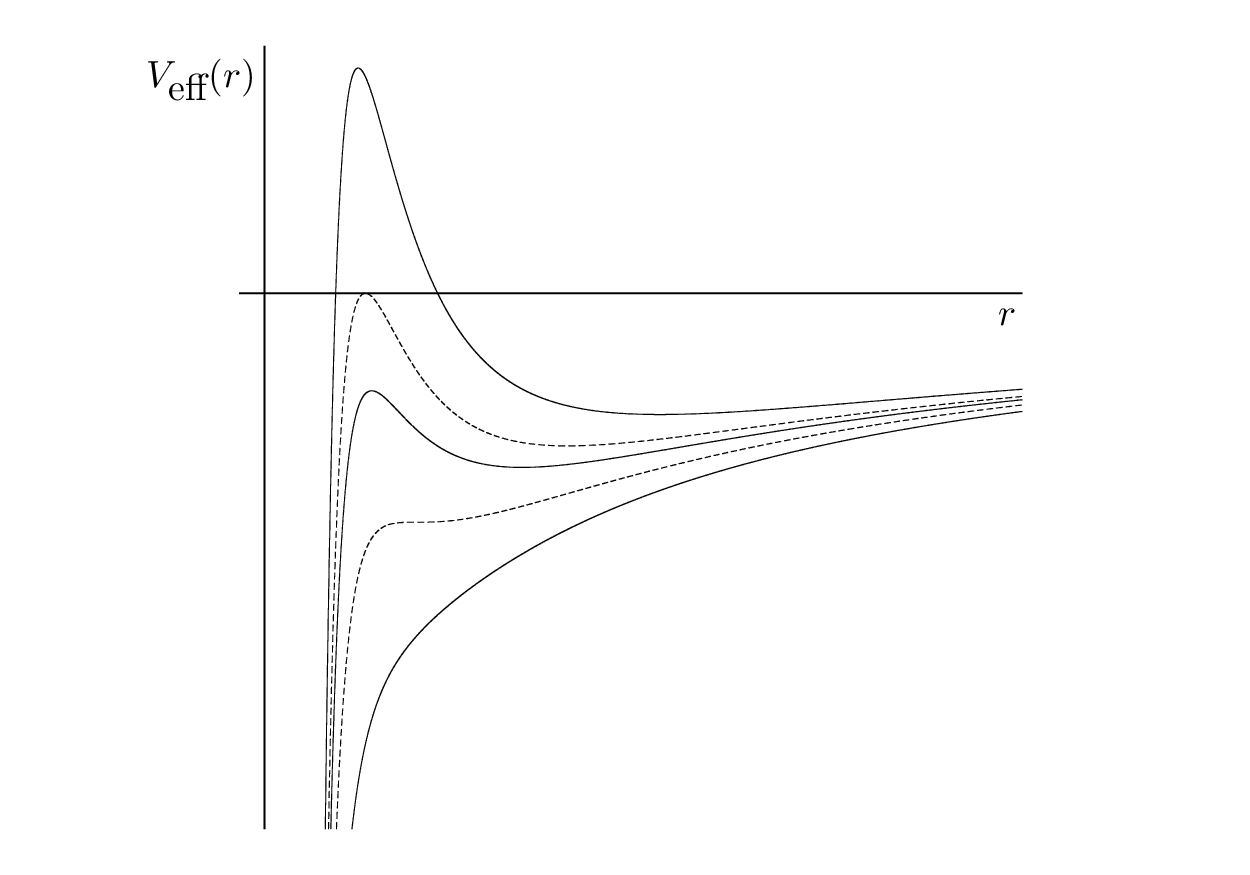}
  \caption{The Schwarzschild effective potential drawn in solid lines
  as a function of radial coordinate $r$ for $L_z = 3, 3.8$ and $4.4$,
  from bottom to top.  The upper and lower dashed lines represent the
  borderline potentials for $L_z = L_{\text{ibco}}$ and $L_z =
  L_{\text{isco}}$, respectively.}
\label{fig:SchVeffallL}
\end{figure}

\begin{figure}
    \includegraphics[width=100mm]{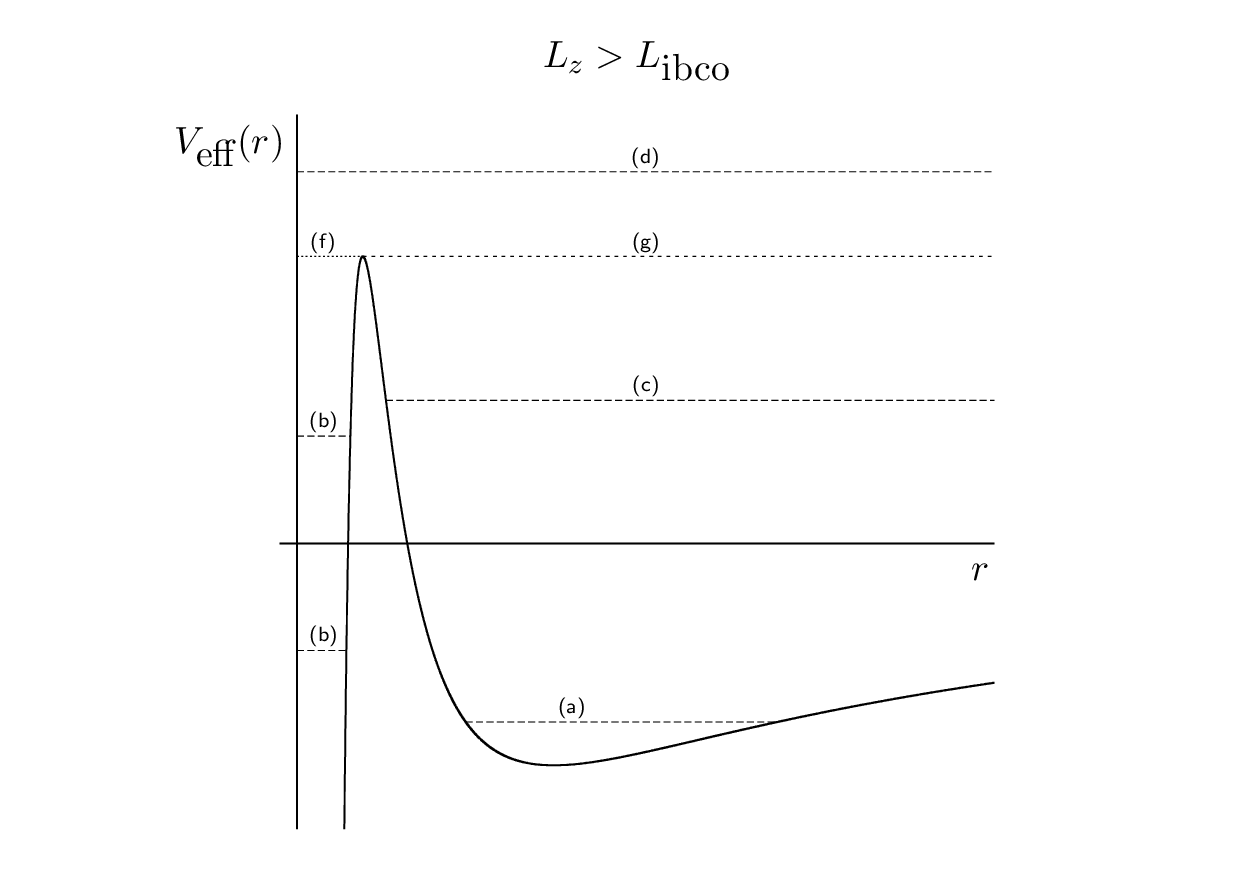}
    \includegraphics[width=100mm]{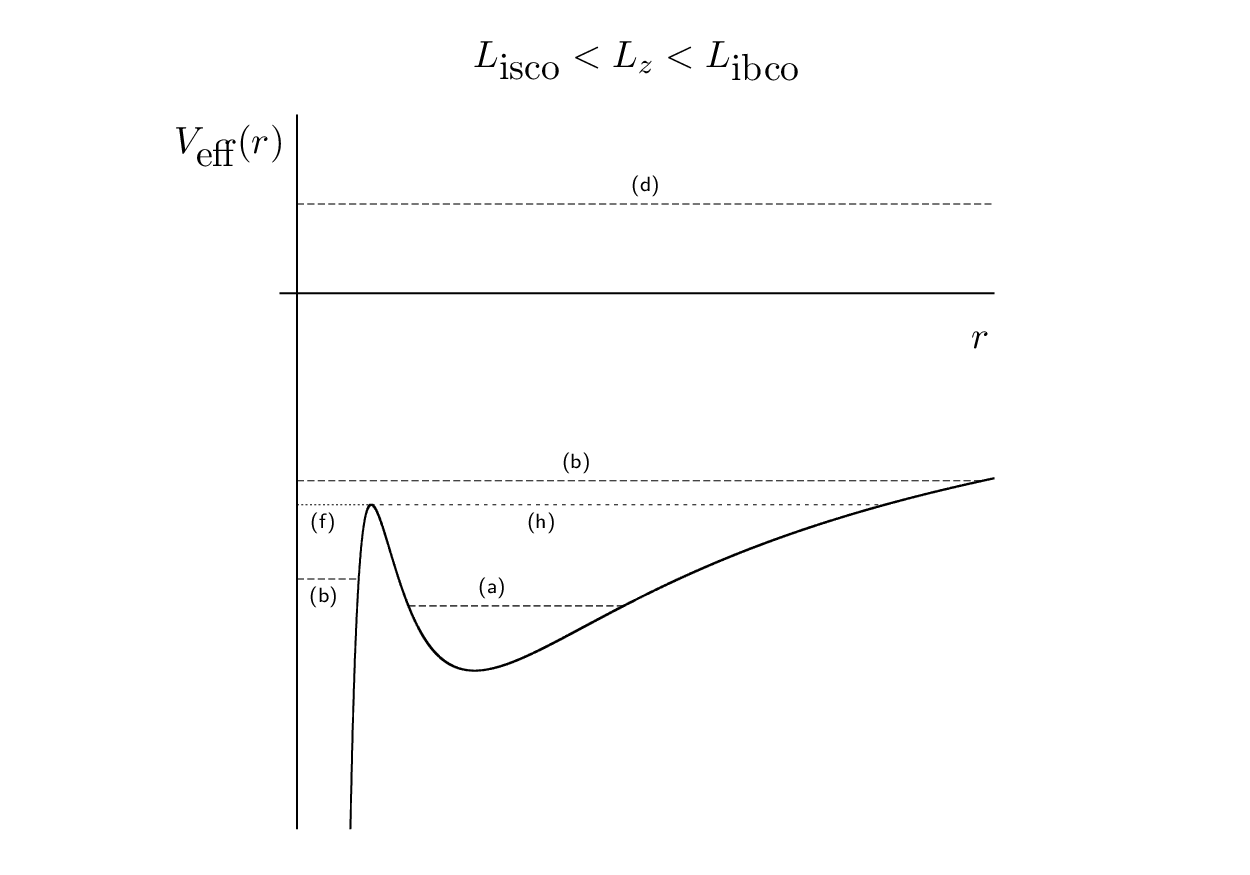}
\hfill
  \caption{Representative plots of the Schwarschild $V_{\text{eff}}(a=0)$
  for a value of $L_z >L_{\text{ibco}}$ (top) and $L_{\text{isco}} <
  L_z < L_{\text{ibco}}$ (bottom).  As explained in the text, the
  horizontal lines define the energies (for the fixed $L_z$ of each
  $V_{\text{eff}}(a=0)$) of orbits that (a) oscillate, (b) plunge, (c)
  escape to $r = \infty$, or (d) both plunge and access $r = \infty$,
  as $t \to \pm\infty $.  The lines (f) and (g) tangent to
  $V_{\text{eff}}(a=0)$ at $r = r_u$ represent $E = E_u$ orbits that
  asymptotically approach $r_u$ at $t = +\infty$ or $-\infty$ (the
  ones marked (f) also plunge).  In the lower figure, the $E = E_u$
  orbit (h) on the right also has a turning point, so it approaches
  $r_u$ at both $t = \pm\infty$ and is homoclinic to the unstable
  circular orbit.}
  \label{fig:SchVeffL>Lisco}
\end{figure}

The value of $L_z$ fixes the form of the potential, as shown in Fig.\
\ref{fig:SchVeffallL}.  Two critical values of $L_z$ define regimes in
which the potential exhibits different qualitative features: the
angular momentum $L_\text{{isco}}$ of the innermost stable circular
orbit (isco), associated with the saddle point in the lower dashed
potential, and the angular momentum $L_{\text{ibco}}$ of the innermost
bound circular orbit (ibco), the circular orbit with $E = 1$.

The effective potential picture allows us to determine the hyperbolic
invariant sets at a glance:  
\begin{itemize}
\item No invariant set exists with $L_z <
L_{\text{isco}}$, since every such orbit plunges and thus fails the
recurrence test.  

\item When $L_z > L_{\text{isco}}$, the potential admits
one stable circular orbit with $r= r_s, E = E_s$ (minimum of
$V_{\text{eff}}(a=0)$) and one unstable circular orbit with $r = r_u, E =
E_u$ (maximum of $V_{\text{eff}}(a=0)$).  

\item The value $L_{\text{ibco}}$ further distinguishes the two
subcases seen in Fig.\ \ref{fig:SchVeffL>Lisco}, from which we see
that orbits with $E \neq E_u$ never approach an invariant set.
Instead, as $t \to \pm\infty$, every such orbit (a) oscillates between
two turning points in the potential well, (b) plunges,\footnote{Eq.\
(\ref{subeq:dimcarter-t}) implies that orbits plunge (reach the
horizon) after a finite amount of proper time $\tau$ but an infinite
amount of coordinate time $t$.} (c) escapes, or (d) escapes as $t \to
-\infty$ and plunges as $t \to +\infty$ (or vice versa).
\end{itemize}
None of the above asymptote to an invariant set.

In contrast, orbits with $E = E_u$ do asymptotically approach an
invariant set, namely the unstable circular orbit.  Consider the upper
panel in Fig.\ \ref{fig:SchVeffL>Lisco}, for which $L_z >
L_{\text{ibco}}$.  There is an $E=E_u$ orbit that approaches $r_u$ as
$t \to -\infty$ and plunges as $t \to +\infty$ and another that
plunges as $t \to -\infty$ and approaches $r_u$ as $t \to +\infty$,
both represented by line (f) in the figure.  These two orbits are
distinct, just as the $E=E_u$ orbit that escapes as $t \to -\infty$
and approaches $r_u$ as $t \to +\infty$ is distinct from its
time-reversed counterpart (both represented by (g)).  So while they
define stable and unstable manifolds for the circular orbit shown,
these $E=E_u$ orbits are non-intersecting (share no initial conditions
$(r, \dot{r})$) and thus are not homoclinic to the circular orbit.

However, when $L_{\text{isco}} < L_z < L_{\text{ibco}}$, as in the
lower panel of Fig.\ \ref{fig:SchVeffallL}, the $E = E_u$ orbit (h)
has a turning point and thus approaches $r_u$ at both $t \to
\pm\infty$.  Parts of the stable and unstable manifolds of this
unstable circular orbit intersect (in fact, they completely coincide),
and these orbits are therefore homoclinic to the circular orbit.

We thus conclude that since they are the only recurrent orbits that
are approached by any other orbit in the infinite future or past, the
unstable circular orbits are the only hyperbolic invariant sets.
Furthermore, those unstable circular orbits with $E < 1$
($L_{\text{isco}} < L_z < L_{\text{ibco}}$) have associated homoclinic
orbits with the same angular momentum and energy, or more
specifically, a family of such orbits differing from one another by an
overall translation in $\varphi$.  References \cite{bombelli1992} and
\cite{levino2000} make similar arguments for the Schwarzschild
case.

\subsection{$a\neq 0$}

Our argument will focus on the roots of the quartic $R$, which we can
rewrite as
\begin{equation}
\label{eq:Rfactored}
  R(r)=(E^2-1)r(r-r_1)(r-r_2)(r-r_3)
  \quad .
\end{equation}
For ease of notation, we adopt the conventions that, from left to right
in (\ref{eq:Rfactored}), real roots appear before complex roots and
the nonzero real roots appear in ascending order $r_1 < r_2 < r_3$.
Additionally,
\begin{alignat}{2}
\label{eq:dRdr0at0}
  R'(r=0) &> 0 &\quad (\text{for } aE \neq L_z)
  \quad ,
\end{alignat}
so $R$ is negative just to the left and positive just to the right of
the root at $r=0$.  Since complex roots occur in conjugate pairs, the
zero root implies that at least one of the three remaining roots is
real.

\begin{figure}[ht]
  \centering
  \hspace{-23mm}
  \begin{minipage}{94mm}
    \begin{center}
      \includegraphics[width=47mm]{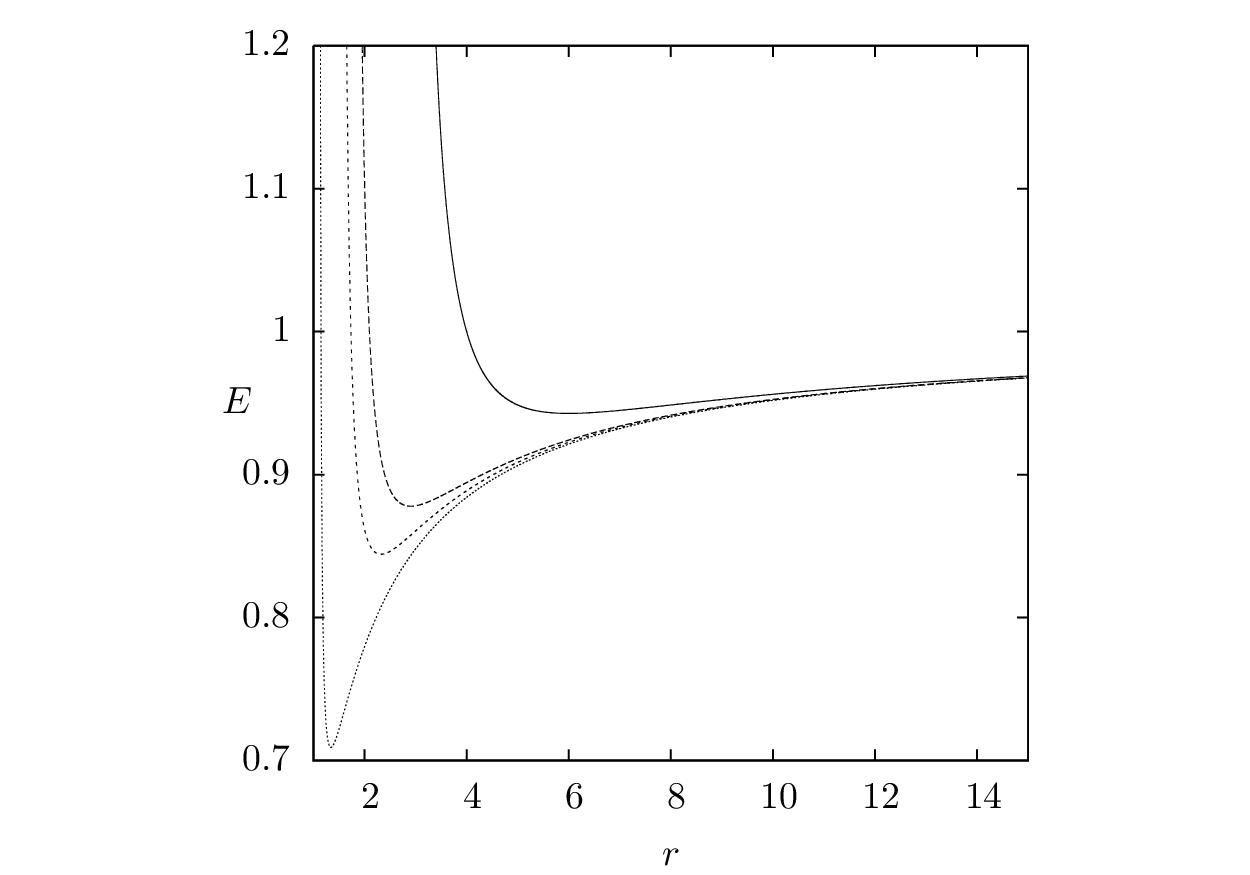}
      \hspace{-6mm}
      \includegraphics[width=47mm]{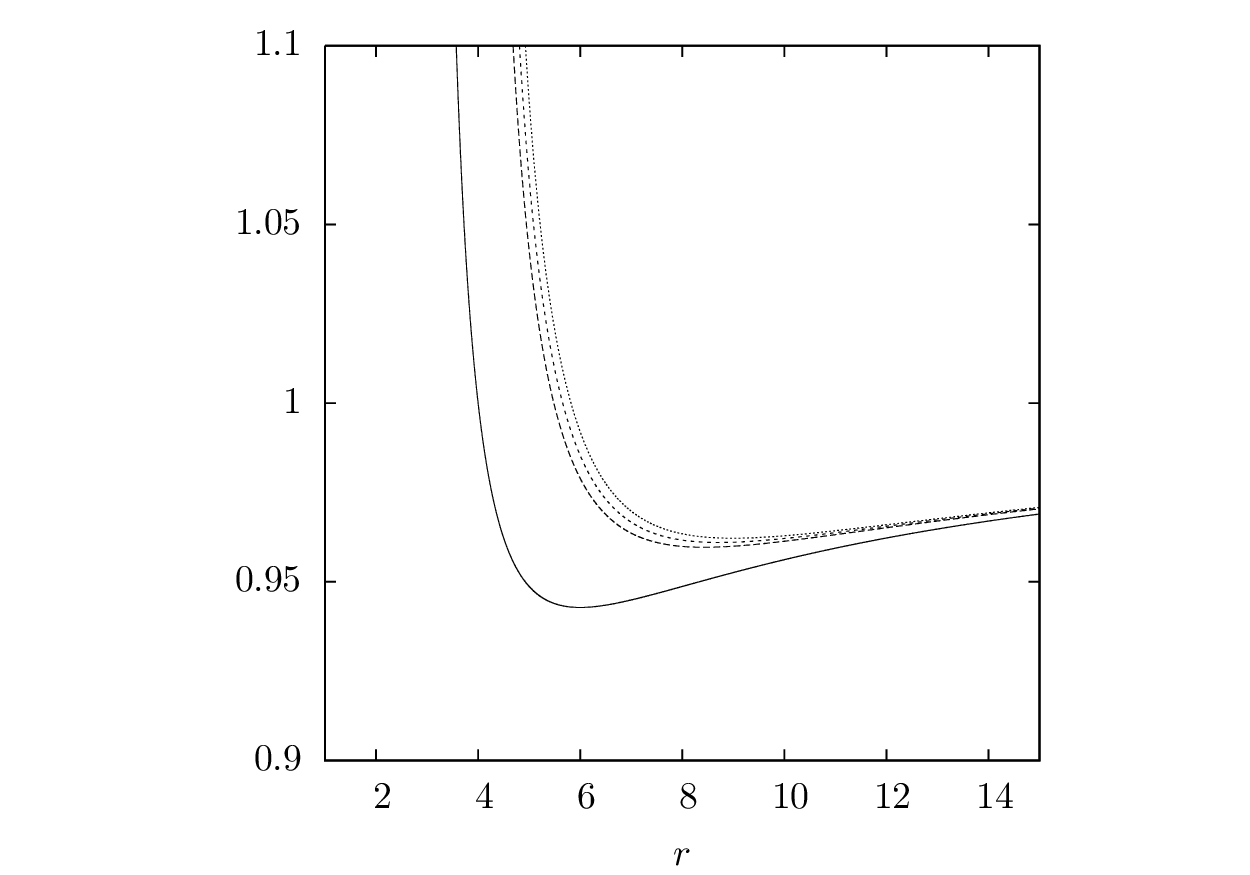}
      \hfill
    \end{center}
  \end{minipage}
  \hfill
  \\
  \hspace{-19mm}
  \begin{minipage}{94mm}
    \begin{center}
      \includegraphics[width=47mm]{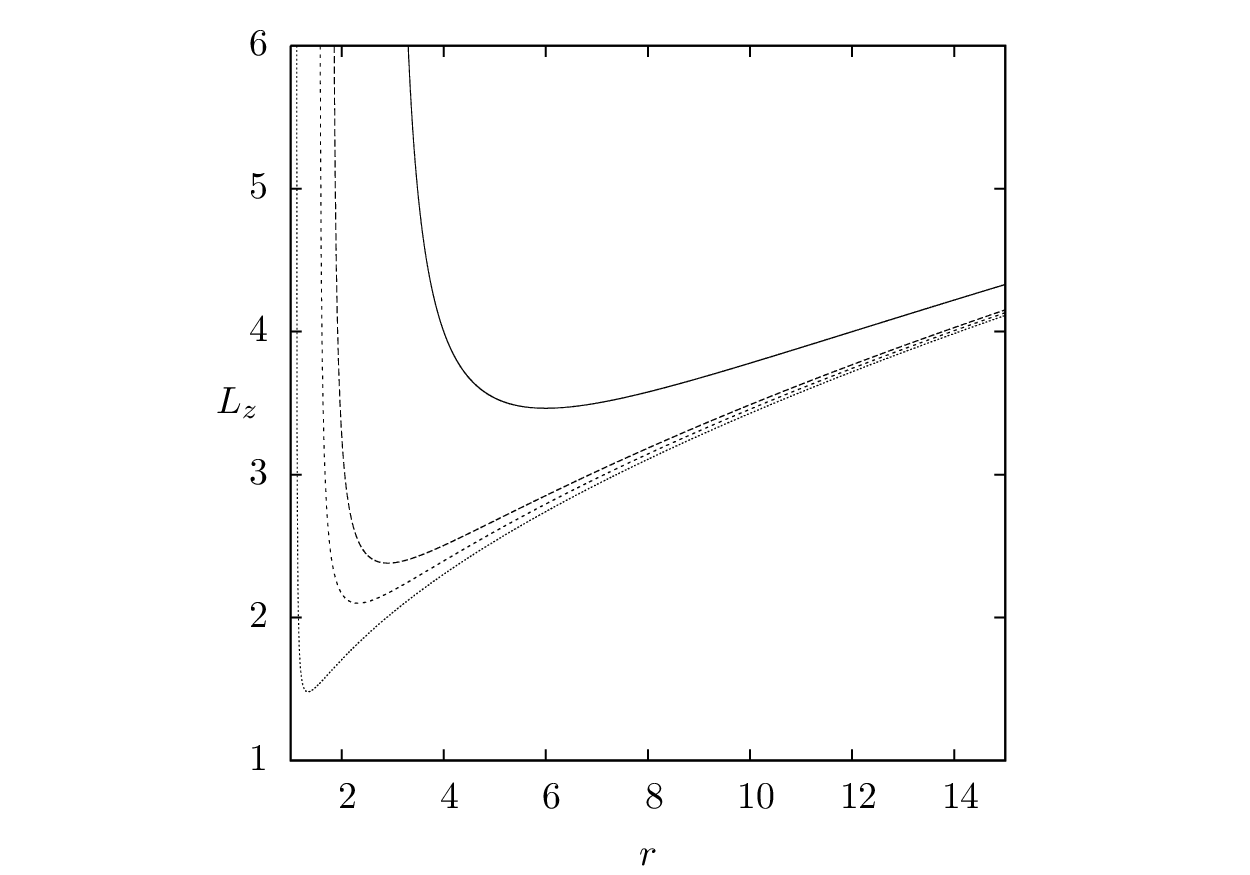}
      \hspace{-5mm}
      \includegraphics[width=47mm]{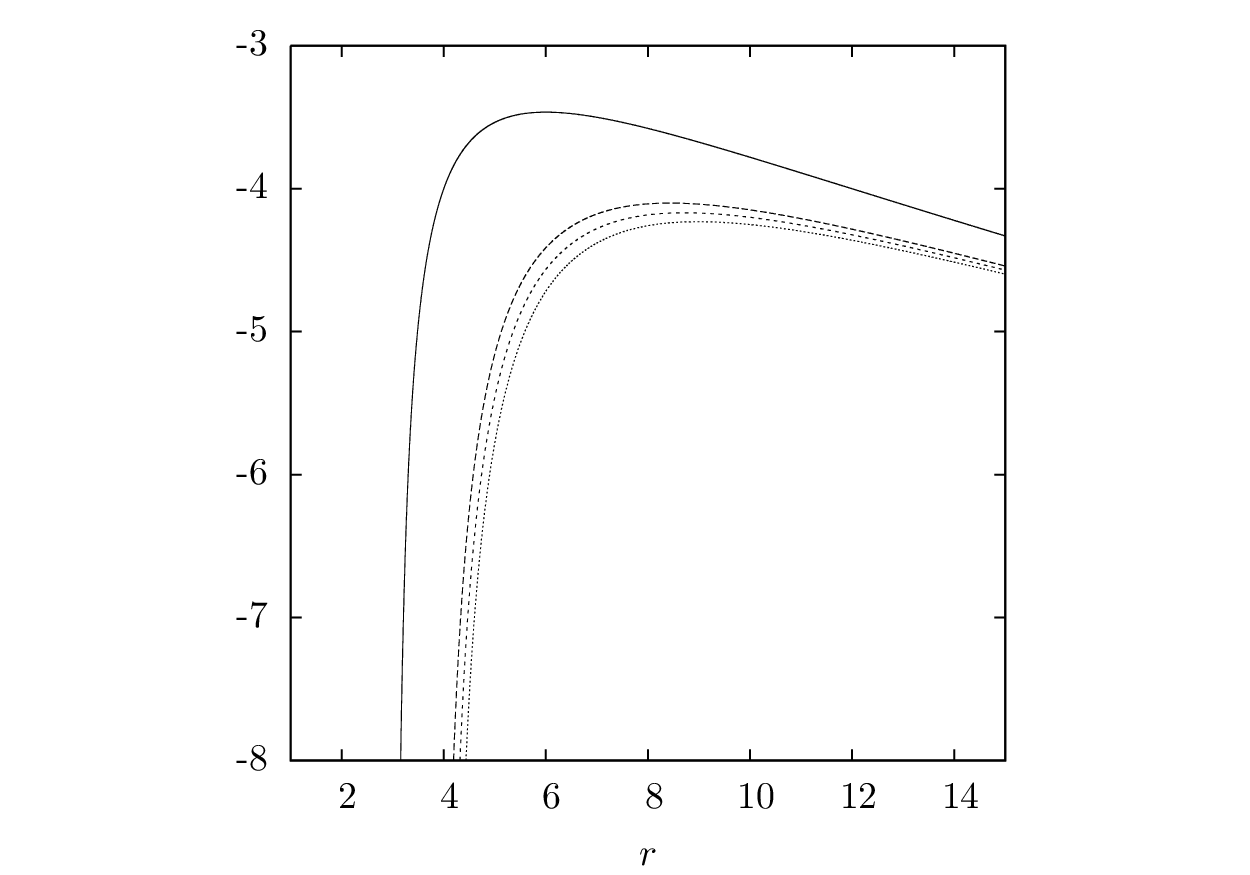}
      \hfill
    \end{center}
  \end{minipage}
  \hfill
  \caption{$E$ and $L_z$ as functions of the radius $r$ of prograde
  (left panels) and retrograde (right panels) circular orbits.  The
  spin parameter for the curves are $a=0$ (solid curve), and then in
  order of increasing distance from the solid curves, $a = 0.8, 0.9$
  and $0.995$.  For a given $a$, $E$ and $L_z$ have simultaneous
  minima at $r = r_{\text{isco}}$.}
  \label{fig:KerrELcirc}
\end{figure}

The non-negativity of $\dot{r}^2$ implies that motion is only possible
where $R(r) \geq 0$.  Turning points of the motion, for which
$V_{\text{eff}}(r) = \varepsilon_{\text{eff}} = 0$, correspond to
single roots of $R(r)$.  Circular orbits require both $V_{\text{eff}}
= 0$ and $\txtD{r}{V_{\text{eff}}} = 0$, or the equivalent
\begin{alignat}{3}
\label{eq:circcond}
  R(r) &= 0 & \quad &\text{and} \quad & R'(r) &= 0
  \quad ,
\end{alignat}
and thus correspond to double roots of $R$, as Fig.\ \ref{RvsVSchFig}
confirms.  Simultaneously solving these equations yields expressions
\cite{Bardeen1972}
\begin{subequations}
\label{app:eq:ELcirc}
\begin{align}
    E &= \ph{\pm} \frac{r^{3/2} - 2r^{1/2} \pm a}
    {r^{3/4}\sqrt{r^{3/2} - 3r^{1/2} \pm 2a}}
    \label{app:eq:Ecirc}\\
    L_z &= \pm \frac{r^2 \mp 2ar^{1/2} + a^2}
    {r^{3/4}\sqrt{r^{3/2} - 3r^{1/2} \pm 2a}}
    \label{app:eq:Lcirc}
\end{align}
\end{subequations}
for the energy and angular momentum of circular orbits, where the
top/bottom signs apply to prograde/retrograde orbits.  These
functions, plotted for a sample of $a$ values in Fig.\
\ref{fig:KerrELcirc} simultaneous minima (maxima for retrograde $L_z$)
at \cite{Bardeen1972}
\begin{alignat}{1}
\label{eq:Kerrisco}
  r_{\text{isco}} &= 3 + Z_2 \mp \sqrt{(3 - Z_1)(3 + Z_1 +2Z_2)} \\
  Z_1 &\equiv 1 + \sqrt[3]{1 - a^2}
  \left[ \sqrt[3]{1 + a} + \sqrt[3]{1 - a} \right] \nonumber \\
  Z_2 &\equiv  \sqrt{3a^2 + Z_1^2} \nonumber
  \quad .
\end{alignat}
Since $R''(r_{\text{isco}}) = 0$ when $E=E_{\text{isco}},
\absval{L_z}=\absval{L_{\text{isco}}}$, the isco corresponds to the
only possible triple root of $R$.

Furthermore, since
\begin{equation}
\label{eq:Vdoubleprime}
  \pdsq{r}{V_{\text{eff}}}
  \evalat{\substack{R\ph{'}=0\,,\\R'=0\ph{\,,}}}
  = -\frac{R''(r)}{2\Sigma^2}
  \quad ,
\end{equation}
$R''$ also determines the stability of circular orbits, with
\begin{alignat}{1}
\label{eq:circstabcond}
  \begin{split}
    r < r_\text{isco} \implies R''(r) &> 0 
    \implies \text{unstable}\\
    r > r_\text{isco} \implies R''(r) &< 0
    \implies \text{stable}
  \end{split}
\end{alignat}
for a given $\absval{L_z} > \absval{L_\text{isco}}$.  Also,
paralleling the $a =0$ case, $E(r)$ in (\ref{app:eq:Ecirc}) increases
monotonically for $r > r_{\text{isco}}$ and approaches 1 as $r \to
\infty$, so that stable circular orbits always have $E_{\text{isco}}<
E < 1$.  Unstable circular orbits, on the other hand, can have any $E
> E_{\text{isco}}$, and the circular orbit with $E = 1$ occurs at
\cite{Bardeen1972}
\begin{equation}
\label{eq:ribcodef}
  r_{\text{ibco}} \equiv 2 \mp a + 2\sqrt{1 \mp a}
  \quad .
\end{equation}

We now show that every non-circular Kerr equatorial orbit falls into
one of the same categories listed for Schwarzschild orbits in Fig.\
\ref{fig:SchVeffL>Lisco}.  Recall that each $R(r)$ plot represents
only those orbits with the same $E$ and $L_z$ and that motion is only
possible in regions where $R > 0$.  When $E > 1$, $R(r) \to +\infty$
at both $r \to \pm\infty$, and (\ref{eq:dRdr0at0}) implies that $R$
has a negative root.  There are thus three possibilites for the number
and type of positive roots:
\begin{itemize}
  \item No positive roots, in which case all positive $r$ are
  accessible, and $R$ represents a single type (d) orbit.
  \item Two positive roots $r_2 < r_3$, resulting in a type (b) orbit
  ($0 \leq r \leq r_2$) and a type (c) orbit ($r \geq r_3$ ).
  \item One positive double root $r_2 = r_3 \equiv r_u$ with $R''(r_u)
  > 0$, in which case $R$ represents an $E_u > 1$ circular orbit at
  $r=r_u$ plus orbits of type (f) ($0 \leq r \leq r_u$) and (g) ($r
  \geq r_u$) that asymptotically approach $r_u$ in either the infinite
  future or past (but not both).
\end{itemize}
The $E = 1$ case is the same as above but without the negative root
(since $R$ is only cubic when $E = 1$).  We thus conclude as in the $a
= 0$ case that the invariant sets with $E \geq 1$ are unstable
circular orbits but that they do not have orbits homoclinic to them.

When $E < 1$, $R \to \infty$ at $r \to \pm\infty$.  Eq.\
(\ref{eq:dRdr0at0}) requires that there be at least one positive root
and only an even number of negative roots, leaving four possibilites
for the number and type of positive roots:
\begin{itemize}
  \item Just the one root $r_1$, resulting in a type (b) orbit ($0
  \leq r \leq r_1$)
  \item Three total positive roots $r_1 < r_2 < r_3$, resulting in a
  type (b) orbit ($0 \leq r \leq r_1$) and an oscillatory bound orbit
  of type (a) ($r_2 \equiv r_p \leq r \leq r_3 \equiv r_a$)
  \item One single positive root $r_1$ and one double root $r_2 \equiv
  r_s > r_1$ with $R''(r_s) < 0$, denoting a type (b) orbit ($0 \leq r
  \leq r_1$) and a stable circular orbit of radius $r_s$
  \item One single root $r_2$ and one double root at $r_1 \equiv r_u <
  r_2$ with $R''(r_u) > 0$, so that $R(r)$ features an unstable
  circular orbit with $E_u < 1$ at $r_u$, a type (b) orbit ($0 \leq r
  \leq r_1$), and a type (h) orbit ($r_u \leq r \leq r_2 \equiv r_a$)
  that approaches $r_u$ as $t \to \pm\infty$, i.e. an orbit homoclinic
  to $r_u$.
\end{itemize}
As before, we conclude that the invariant sets with homoclinic orbits
are the unstable circular orbits with $E_u < 1$, one of which exists
for every $r_{\text{ibco}} < r_u < r_{\text{isco}}$.

\vfill\eject

\section{Derivation of Equatorial Homoclinic Orbits}
\label{append1}

We now find an exact solution by analytically integrating the
following equations:
\begin{align}
  \tau(r) &= -\int^r_{r_0} dr \D{r}{\tau} = 
  -\int^r_{r_0} dr \frac{\Sigma}{\sqrt{R}}
  \label{eq:taureq}\\
  \begin{split}
    t(r) &= -\int^r_{r_0} dr\,
    \frac{\txtD{\tau}{t}}{\txtD{\tau}{r}}\\
    &= -\int^r_{r_0} dr\,
    \frac{r^2(r^2 + a^2)E + 2a(aE - L_z)r}{\Delta\sqrt{R}} \quad ,
  \end{split}
  \label{eq:treq}\\
  \begin{split}
    \varphi(r) &= -\int^r_{r_0} dr\,
    \frac{\txtD{\tau}{\varphi}}{\txtD{\tau}{r}}\\
    &= -\int^r_{r_0} dr\,
    \frac{r^2L + 2(aE - L_z)r}{\Delta\sqrt{R}}
  \end{split}
  \label{eq:phireq}
\end{align}
where $r$ and $r_0$ are both radial coordinates along the same phase
(i.e. along a given half-leaf) of the motion.  Removing the overall
minus signs yields the corresponding expressions for outbound motion.

\subsection{Integral equations for homoclinic orbits}
Some algebraic manipulation of the denominators of the integrals
\eqref{eq:taureq}-\eqref{eq:phireq} renders them more suitable for
evaluation.  We begin with $R(r)$, which for equatorial orbits has
its smallest root at $r = 0$ (since $Q = 0$).  For homoclinic orbits
specifically, the remaining roots of $R$ are a double root at $r_u (=
r_p)$ and a simple root at $r_a$.  $R(r)$ therefore factors into
\begin{align}
  R(r) &= (E^2 - 1) (r - r_u)^2 r (r - r_a)\\
  &= (1 - E^2) (r - r_u)^2 r (r_a - r) \, ,
\label{eq:Reqfactoredr}
\end{align}
where we've written $R$ in the second form so that the product of the
$r$-dependent terms is manifestly positive for $r_p < r < r_a$.  The
square root in the denominators thus becomes
\begin{equation}
  \sqrt{R(r)} = \sqrt{1 - E^2}(r - r_u) \sqrt{r(r_a - r)} \quad ,
\end{equation}
where we have replaced $\sqrt{(r - r_u)^2}$ with $(r - r_u)$ since $r
> r_u$ over the entire orbit.  $\Delta$ also factors into
\begin{equation}
  \Delta = (r - r_{\sss +})(r - r_{\sss -}) \quad ,
\label{eq:Deltafactored}
\end{equation}
where $r_{\sss +}\equiv 1+\sqrt{1 - a^2}$ and $r_{\sss -}\equiv
1-\sqrt{1 - a^2}$ are the outer and inner horizons, respectively, of
the central black hole.  The integrals
\eqref{eq:taureq}-\eqref{eq:phireq} are therefore
\begin{align}
  \begin{split}
    \tau(r) &= -\frac{1}{\sqrt{1 - E^2}} \times {}\\
    &\mrph{=} \nqquad \int^r_{r_0} dr\,
    \frac{r^2}
	 {
	   \ph{(r - r_{\sss +})}
	   (r - r_u)\sqrt{r (r_a - r)}
	   \ph{(r - r_{\sss -})}
	 }
  \end{split}
  \label{eq:tauhomofactoredr}\\
  \begin{split}
    t(r) &= -\frac{1}{\sqrt{1 - E^2}} \times {}\\
    &\mrph{=} \nqquad \int^r_{r_0} dr\,
    \frac{r^2(r^2 + a^2)E + 2a(aE - L_z)r}
	 {
	   (r - r_{\sss +})(r - r_{\sss -})(r - r_u)\sqrt{r (r_a - r)}
	 }
  \end{split}
  \quad .
  \label{eq:thomofactoredr}\\
  \begin{split}
    \varphi(r) &= -\frac{1}{\sqrt{1 - E^2}} \times {}\\
    &\mrph{=} \nqquad \int^r_{r_0} dr\,
    \frac{r^2L_z + 2(aE - L_z)r}
	 {
	   (r - r_{\sss +})(r - r_{\sss -})(r - r_u)\sqrt{r (r_a - r)}
	 }
  \end{split}
  \label{eq:phihomofactoredr}
\end{align}

\subsection{Change of variable}
We can express the integrals above more compactly as
\begin{align}
  \tau(r) &=
  \frac{1}{\sqrt{1 - E^2}}\,
  I_1
  \label{eq:tauhomoIs}\\
  \begin{split}
    t(r) &=
    \frac{1}{\sqrt{1 - E^2}} \times {}\\
    &\mrph{=}
    \lf[
      E I_2 + a^2 E I_3 + 2a(aE - L_z) I_4
      \rt]
  \end{split}
  \quad ,
  \label{eq:thomoIs}\\
  \varphi(r) &=
  \frac{1}{\sqrt{1 - E^2}}
  \lf[
    L_z I_3 + 2(aE - L_z) I_4
    \rt]
  \label{eq:phihomoIs}
\end{align}
where
\begin{subequations}
\label{eq:Iints}
  \begin{align}
    I_1 &\equiv  
    -\int^r_{r_0} dr \, \frac{r^2}
    {\ph{(r - r_{\sss +})}(r - r_u)\sqrt{r (r_a - r)}\ph{(r - r_{\sss -})}}
    \label{eq:I1int}\\  
    I_2 &\equiv
    -\int^r_{r_0} dr \, \frac{r^4}
    {(r - r_{\sss +})(r - r_{\sss -})(r - r_u)\sqrt{r (r_a - r)}}
    \label{eq:I2int}\\  
    I_3 &\equiv
    -\int^r_{r_0} dr \, \frac{r^2}
    {(r - r_{\sss +})(r - r_{\sss -})(r - r_u)\sqrt{r (r_a - r)}}
    \elpunc{.}
    \label{eq:I3int}\\  
    I_4 &\equiv
    -\int^r_{r_0} dr \, \frac{r}
    {(r - r_{\sss +})(r - r_{\sss -})(r - r_u)\sqrt{r (r_a - r)}}
    \label{eq:I4int}
\end{align}
\end{subequations}
We evaluate each of the integrals in \eqref{eq:Iints} in closed form
by the same procedure.  First, we bring one (positive definite) $r$
from each numerator under a radical as an $r^2$:
\begin{subequations}
\label{eq:Iintsrad}
  \begin{align}  
    I_1 &=  
    -\int^r_{r_0} dr \, \sqrt{\frac{r}{r_a - r}}
    \frac{r}{\ph{(r - r_{\sss +})}(r - r_u)\ph{(r - r_{\sss -})}}
    \label{eq:I1intrad}\\  
    I_2 &=
    -\int^r_{r_0} dr \, \sqrt{\frac{r}{r_a - r}}
    \frac{r^3}{(r - r_{\sss +})(r - r_{\sss -})(r - r_u)}
    \label{eq:I2intrad}\\  
    I_3 &=
    -\int^r_{r_0} dr \, \sqrt{\frac{r}{r_a - r}}
    \frac{r}{(r - r_{\sss +})(r - r_{\sss -})(r - r_u)}
    \elpunc{.}
    \label{eq:I3intrad}\\  
    I_4 &=
    -\int^r_{r_0} dr \, \sqrt{\frac{r}{r_a - r}}
    \frac{1}{(r - r_{\sss +})(r - r_{\sss -})(r - r_u)}
    \label{eq:I4intrad}
\end{align}
\end{subequations}
Next, the change of variable
\begin{equation}
\begin{split}
  u = \sqrt{\frac{r_a - r}{r}}\,, \ph{m} r = \frac{r_a}{u^2 + 1}\\
  -dr\,\sqrt{\frac{r}{r_a - r}} = du\, \frac{2 r_a}{(u^2 + 1)^2}
\end{split}
\end{equation}
turns the integrals \eqref{eq:Iintsrad} into
\begin{subequations}
\label{eq:Iintsu}
  \begin{align}  
    I_1 &=  
    2 r_a^2 \int^u_{u_0} du \,
    \frac{1}{\ph{\lf(r_a - r_{\sss +}y\rt)}
      y^2
      \lf(r_a - r_u y\rt)
      \ph{\lf(r_a - r_{\sss -}y\rt)}}
    \label{eq:I1intu}\\  
    I_2 &=
    2 r_a^4 \int^u_{u_0} du \,
    \frac{1}{y^2
      \lf(r_a - r_{\sss +}y\rt)
      \lf(r_a - r_{\sss -}y\rt)
      \lf(r_a - r_u y\rt)}
    \label{eq:I2intu}\\  
    I_3 &=
    2 r_a^2 \int^u_{u_0} du \,
    \frac{1}{\lf(r_a - r_{\sss +}y\rt)
      \lf(r_a - r_{\sss -}y\rt)
      \lf(r_a - r_u y\rt)}
    \elpunc{,}
    \label{eq:I3intu}\\  
    I_4 &=
    2 r_a^{\ph{2}} \int^u_{u_0} du \,
    \frac{y}{\lf(r_a - r_{\sss +}y\rt)
      \lf(r_a - r_{\sss -}y\rt)
      \lf(r_a - r_u y\rt)}
    \label{eq:I4intu}
\end{align}
\end{subequations}
where we've written $y \equiv u^2 + 1$ as a shorthand.

\subsection{Partial fraction decomposition}
Each integrand in \eqref{eq:Iintsu} is now a product of factors linear
in $y$ and splits up via a standard partial fraction decomposition.
We get
\begin{subequations}
\label{eq:IintsPF}
  \begin{align}  
    I_1 &=  
    2 \int^u_{u_0} du \,\lf[
      \ph{\frac{1}{y}} \frac{A_{11}}{y^2} \ph{\frac{1}{y}}
      + \ph{\!m} \frac{A_{12}}{y} \ph{\!m}
      + \ph{\frac{1}{y}} \frac{A_{13}}{r_a - r_u y} \ph{\frac{1}{y}}
      \rt]
    \label{eq:I1intPF}\\  
    \begin{split}
      I_2 &=
      2 \int^u_{u_0} du \, \lf[
	\frac{A_{21}}{y^2} + \frac{A_{22}}{y}
	+ \frac{A_{23}}{r_a - r_u y}
	\rt.\\
	&\ph{{} = 2 \int^u_{u_0} du \,} \lf.
	      \ph{\frac{A_{23}}{r_a - r_u y}\,\,}
		 {}+ \frac{A_{24}}{r_a - r_{\sss +}y}
		 + \frac{A_{25}}{r_a - r_{\sss -}y}
	      \rt]
    \end{split}
    \label{eq:I2intPF}\\  
    I_3 &=
    2 \int^u_{u_0} du \, \lf[
    \frac{A_{33}}{r_a - r_u y}
    + \frac{A_{34}}{r_a - r_{\sss +}y}
    + \frac{A_{35}}{r_a - r_{\sss -}y}
    \rt]
    \elpunc{,}
    \label{eq:I3intPF}\\  
    I_4 &=
    2 \int^u_{u_0} du \, \lf[
    \frac{A_{43}}{r_a - r_u y}
    + \frac{A_{44}}{r_a - r_{\sss +}y}
    + \frac{A_{45}}{r_a - r_{\sss -}y}
    \rt]
    \label{eq:I4intPF}
\end{align}
\end{subequations}
where
\begin{widetext}
\begin{alignat}{3}
  A_{11} ={} &r_a = A_{21}\quad
  & A_{12} &= r_u \quad
  & A_{22}=r_u+r_{+}+r_{-}&\quad A_{13} = r_u^2
  \nonumber\\
  A_{23} &= r_u^3 A_{43}
  & &A_{33} = r_u A_{43}
  & A_{43} &= \frac{r_u}
  {\lf(r_u - r_{\sss +}\rt)\lf(r_u - r_{\sss -}\rt)}
  \nonumber\\
  A_{24} &= r_{\sss +}^3 A_{44}
  & &A_{34} = r_{\sss +} A_{44}  
  & A_{44} &= \frac{-r_{\sss +}}
  {\lf(r_{\sss +} - r_{\sss -}\rt)\lf(r_u - r_{\sss +}\rt)}
  \elpunc{.}
  \label{eq:As}\\
  A_{25} &= r_{\sss -}^3 A_{45}
  & &A_{35} = r_{\sss -} A_{45}  
  & A_{45} &= \frac{r_{\sss -}}
  {\lf(r_{\sss +} - r_{\sss -}\rt)\lf(r_u - r_{\sss -}\rt)}
  \nonumber
\end{alignat}
\end{widetext}
We are left with five different integrals to calculate.  Recalling
that $y = u^2 + 1$, those antiderivatives evaluate to
\begin{subequations}
\label{eq:5uints}
\begin{align}
  \begin{split}
    \mathcal{I}_1 &\equiv \int du \, \frac{1}{y^2}
    = \int du \, \frac{1}{\lf(u^2 + 1\rt)^2}\\
    &= \frac{1}{2} \lf(\frac{u}{1 + u^2} + \atan u\rt)
  \end{split}
  \label{eq:ysqint}\\
  \begin{split}
    \mathcal{I}_2 &\equiv \int du \, \frac{1}{y}
    = \int du \, \frac{1}{u^2 + 1}\\
    &= \atan u
  \end{split}
  \label{eq:yint}\\
  \begin{split}
    \mathcal{I}_3 &\equiv \int du \, \frac{1}{r_a - r_u y}
    = \int du \, \frac{1}{\lf(r_a - r_u\rt) - r_u u^2 }\\
    &= \frac{1}{\sqrt{r_u}\sqrt{r_a - r_u}}
    \atanh\lf\{u \sqrt{\frac{r_u}{r_a - r_u}}\rt\}
    \elpunc{.}
  \end{split}
  \label{eq:ruyint}\\
  \begin{split}
    \mathcal{I}_4 &\equiv \int du \, \frac{1}{r_a - r_{\sss +} y}
    = \int du \, \frac{1}{\lf(r_a - r_{\sss +}\rt) - r_{\sss +} u^2 }\\
    &= \frac{1}{\sqrt{r_{\sss +}}\sqrt{r_a - r_{\sss +}}}
    \atanh\lf\{u \sqrt{\frac{r_{\sss +}}{r_a - r_{\sss +}}}\rt\}
  \end{split}
  \label{eq:rplsyint}\\
  \begin{split}
    \mathcal{I}_5 &\equiv \int du \, \frac{1}{r_a - r_{\sss -} y}
    = \int du \, \frac{1}{\lf(r_a - r_{\sss -}\rt) - r_{\sss -} u^2 }\\
    &= \frac{1}{\sqrt{r_{\sss -}}\sqrt{r_a - r_{\sss -}}}
    \atanh\lf\{u \sqrt{\frac{r_{\sss -}}{r_a - r_{\sss -}}}\rt\}
  \end{split}
  \label{eq:rmnsyint}
\end{align}
\end{subequations}

The integrals \eqref{eq:IintsPF} are therefore
\begin{subequations}
\label{eq:IswithAs}
\begin{align}
  I_1 &= 2 \lf( A_{11}\mathcal{I}_1 + A_{12}\mathcal{I}_2
    + A_{13}\mathcal{I}_3 \rt)\\
  \begin{split}
    I_2 &= 2 \lf( A_{21}\mathcal{I}_1 + A_{22}\mathcal{I}_2
      + A_{23}\mathcal{I}_3\rt.\\
      &\quad \lf.
      {} + A_{24}\mathcal{I}_4 + A_{25}\mathcal{I}_5
      \rt)
      \qquad\qquad\qquad .
  \end{split}\\
  I_3 &= 2 \lf( A_{33}\mathcal{I}_3 + A_{34}\mathcal{I}_4
    + A_{35}\mathcal{I}_5 \rt)\\
  I_4 &= 2 \lf( A_{43}\mathcal{I}_3 + A_{44}\mathcal{I}_4
    + A_{45}\mathcal{I}_5 \rt)
\end{align}
\end{subequations}

Combining equations \eqref{eq:tauhomoIs} - \eqref{eq:phihomoIs},
\eqref{eq:As}, \eqref{eq:5uints} and \eqref{eq:IswithAs}, and
recalling that $u = \sqrt{(r_a - r) / r}$, the dynamical variables
$\tau, t$ and $\varphi$ become
\begin{align}
  \tau(r) &= \frac{1}{\sqrt{1 - E^2}}
  \sum_{j=1}^{3} C_j^{(\tau)} f_j(r)
  \label{eq:tauhomoCfs}\\
  t(r) &= \frac{1}{\sqrt{1 - E^2}}
  \sum_{j=1}^{5} C_j^{(t)} f_j(r)
  \qquad ,
  \label{eq:thomoCfs}\\
  \varphi(r) &= \frac{1}{\sqrt{1 - E^2}}
  \sum_{j=3}^{5} C_j^{(\varphi)} f_j(r)
  \label{eq:phihomoCfs}
\end{align}
where the functions $f_j(r)$ are
\begin{subequations}
\label{eq:fs}
\begin{align}
  f_1(r) &= \sqrt{r\lf(r_a - r\rt)}
  \label{eq:f1}\\
  f_2(r) &= \atan \sqrt{\frac{r_a - r}{r}}
  \label{eq:f2}\\
  f_3(r) &= \atanh
  \sqrt{
    \frac{r_u}{r_a - r_u}
    \frac{r_a - r}{r}
  }
  \label{eq:f3}\\
  f_4(r) &= \atanh
  \sqrt{
    \frac{r_{\sss +}}{r_a - r_{\sss +}}
    \frac{r_a - r}{r}
  }
  \label{eq:f4}\\
  f_5(r) &= \atanh
  \sqrt{
    \frac{r_{\sss -}}{r_a - r_{\sss -}}
    \frac{r_a - r}{r}
  }
  \label{eq:f5}
\end{align}
\end{subequations}
and the corresponding coefficients are
\begin{widetext}
\begin{alignat}{2}
  C_1^{(\tau)} &= 1 \qquad\quad\,\,
  C_2^{(\tau)} = \lf(r_a + 2 r_u\rt)
  & C_3^{(\tau)} &= 2\sqrt{\frac{r_u^3}{r_a - r_u}}
  \nonumber\\
  C_1^{(t)} &= E
  & C_2^{(t)} &= E\lf(r_a + 2 (r_u + 2)\rt)
  \nonumber\\
  \begin{split}
    C_3^{(t)} &= 2\sqrt{\frac{r_u^3}{r_a - r_u}} \,\times\\
    &\mrph{=} \frac{r_u^2 \lf(r_u^2 + a^2\rt) E
      + 2 a \lf(aE - L_z\rt) r_u}
    {\lf(r_u - r_{\sss +}\rt)\lf(r_u - r_{\sss -}\rt) r_u^2}
  \end{split} &
  \begin{split}
    C_3^{(\varphi)} &= 2\sqrt{\frac{r_u^3}{r_a - r_u}} \,\times\\
    &\mrph{=} \frac{r_u^2 L_z + 2 \lf(aE - L_z\rt) r_u}
    {\lf(r_u - r_{\sss +}\rt)\lf(r_u - r_{\sss -}\rt) r_u^2}
    \qquad\quad .
  \end{split}
  \label{eq:Cs}\\
  \begin{split}
    C_4^{(t)} &= \frac{-4 r_{\sss +}}{r_{\sss +} - r_{\sss -}}
    \,\times\\
    &\mrph{=} \frac{2 E r_{\sss +} - aL_z}
	{\sqrt{r_{\sss +} \lf(r_a - r_{\sss +}\rt)}
	  \lf(r_u - r_{\sss +}\rt)}
  \end{split} &
  \begin{split}
    C_4^{(\varphi)} &= \frac{-2 r_{\sss +}}{r_{\sss +} - r_{\sss -}}
    \,\times\\
    &\mrph{=} \frac{2aE - L_zr_{\sss -}}
	{\sqrt{r_{\sss +} \lf(r_a - r_{\sss +}\rt)}
	  \lf(r_u - r_{\sss +}\rt)}	
  \end{split}
  \nonumber\\
  \begin{split}
    C_5^{(t)} &= \frac{4 r_{\sss -}}{r_{\sss +} - r_{\sss -}}
    \,\times\\
    &\mrph{=} \frac{2 E r_{\sss -} - aL_z}
	{\sqrt{r_{\sss -} \lf(r_a - r_{\sss -}\rt)}
	  \lf(r_u - r_{\sss -}\rt)}
  \end{split} &
  \begin{split}
    C_5^{(\varphi)} &= \frac{2 r_{\sss -}}{r_{\sss +} - r_{\sss -}}
    \,\times\\
    &\mrph{=} \frac{2aE - L_zr_{\sss +}}
	{\sqrt{r_{\sss -} \lf(r_a - r_{\sss -}\rt)}
	  \lf(r_u - r_{\sss -}\rt)}
  \end{split}
  \nonumber
\end{alignat}
\end{widetext}
In the expressions above, we have used the facts that (since $r_{\sss
\pm}$ are roots of $\Delta$) $r_{\sss \pm}^2 + a^2 = 2r_{\sss \pm}$
and that $r_{\sss +} + r_{\sss -} = 2$.

Note that since all of the $f_j(r)$ vanish at $r = r_a$, our
expressions \eqref{eq:tauhomoCfs} - \eqref{eq:phihomoCfs} implicitly
assume the natural choice of time and azimuthal origins, namely at
apastron.  After all, since they are single-leaf orbits with formally
infinite radial periods, homoclinic orbits have only 1 outbound and 1
inbound phase each and transit through apastron only once.  We now
make that choice explicit.  From here on, all expressions assume that
\begin{equation}
\tau(r_a) = t(r_a) = \varphi(r_a) = 0 \quad ,
\label{eq:origin}
\end{equation}
along homoclinic orbits, so that $\tau$ and $t$ are positive/negative
along the inbound/outbound branch, while $\varphi$ is
positive/negative along the inbound/outbound branch for prograde
orbits (increasing $\varphi$) and negative/positive along the
inbound/outbound branch for retrograde orbits (decreasing $\varphi$).

\subsection{Simplification of coefficients}
The task now is to render the coefficients in a more meaningful form.
The $C_1$'s are already simple.  To simplify the $C_2$'s, we expand
the factored form (\ref{eq:Reqfactoredr}) of $R(r)$to
\begin{equation}
  \begin{split}
    R(r) &= (1 - E^2) \times\\
    &\mrph{=} \lf\{-r^4 + (2r_u + r_a)r^3
    - r_u(2r_a + r_u)r^2 + r_u^2 r_a r\rt\}
  \end{split}
  \quad .
\label{eq:Reqfactoredexpanded}
\end{equation}
Comparing to \eqref{eq:Rpoly} (with $Q = 0$) and equating coefficients
of corresponding powers of $r$, we see that for equatorial homoclinic
orbits,
\begin{equation}
  r_a + 2r_u = \frac{2}{1 - E^2} \quad .
\label{eq:2ruplusra}
\end{equation}
Thus, the $C_2$'s are
\begin{align}
  C_2^{(\tau)} &= \frac{2}{1 - E^2}\\
  \begin{split}
    C_2^{(t)} &= E \lf(\frac{2}{1 - E^2} + 4\rt)
    \qquad .\\
    &= 2 E \lf(\frac{3 - 2 E^2}{1 - E^2}\rt)
  \end{split}
\end{align}

For the $C_3$'s, notice that
\begin{align}
  \begin{split}
    R''(r) &= (1 - E^2) \times{}\\
    &\mrph{=} \lf\{-12r^2 + 6(2r_u + r_a)r -2r_u(2r_a + r_u) \rt\}
  \end{split}\\
    \implies R''(r_u) &= (1 - E_u^2) 2r_u(r_a - r_u)
  \quad .
\label{eq:R''}
\end{align}
Inserting this into the expression for the proper time stability
exponent $\gamma\lambda_r$ of the unstable circular orbit associated
with the homoclinic orbit yields
\begin{equation}
  \begin{split}
    \gamma\lambda_r &= \sqrt{\frac{R''(r_u)}{2\Sigma_u^2}}\\
    &= \sqrt{1 - E_u^2}
    \sqrt{\frac{2r_u(r_a - r_u)}{2r_u^4}}\\
    &= \sqrt{1 - E_u^2} \sqrt{\frac{r_a - r_u}{r_u^3}}
  \end{split}\quad .
\label{eq:lamtau}
\end{equation}
The coefficient $C_3^{(\tau)}$ is therefore
\begin{equation}
  C_3^{(\tau)} = 2\sqrt{\frac{r_u^3}{r_a - r_u}}
  = \frac{2}{\gamma\lambda_r} \sqrt{1 - E_u^2} \quad .
\end{equation}
$C_3^{(t)}$ and $C_3^{(\varphi)}$ are each $C_3^{(\tau)}$ times
another factor.  Comparing to \eqref{eq:taureq} and \eqref{eq:phireq},
however, we can identify these extra factors as the $\txtD{\tau}{t}$
and $\txtD{\tau}{\varphi}$, respectively, of the unstable circular
orbit associated with the homoclinic orbit.  That allows us to write
\begin{equation}
  C_3^{(t)} = \frac{2}{\gamma\lambda_r}
  \sqrt{1 - E^2} \D{\tau}{t}(r_u)
  = \frac{2}{\lambda_r}\sqrt{1 - E^2} \quad ,
\end{equation}
where $\lambda_r$, as we will show in paper II \cite{perez-giz2008},  
refers to the stability exponent
governing the evolution with respect to coordinate time $t$ of small
perturbations to the circular orbit (we've used here the fact that
$\lambda_r\,dt = \gamma\lambda_r d\tau$).  Likewise,
\begin{align}
  \begin{split}
    C_3^{(\varphi)} &= \frac{2}{\gamma\lambda_r} 
    \sqrt{1 - E^2} \D{\tau}{\varphi}(r_u)
    = \frac{2}{\lambda_r}\D{t}{\tau}(r_u) \D{\tau}{\varphi}(r_u)\\
    &= \frac{2}{\lambda_r} \sqrt{1 - E^2} \D{t}{\varphi}(r_u)
    = 2\frac{\Omega_u}{\lambda_r} \sqrt{1 - E^2}
  \end{split}
  \quad ,
\end{align}
where
$\Omega_u\equiv \frac{d\varphi}{dt}(r_u)$.

Simplifying the $C_4$'s take a little more work.  As mentioned in \S
~\ref{physspace}, the energy and angular momentum of the homoclinic
orbit are the same as those of the unstable circular orbit at $r_u$.
Recalling the expressions (\ref{app:eq:Lcirc}) for circular orbits
(top/bottom signs are for prograde/retrograde orbits) from
\cite{Bardeen1972}, we can rewrite the numerator of the second factor
in $C_4^{(t)}$ as
\begin{widetext}
\begin{align}
  \begin{split}
    2Er_{\sss +} - aL_z &=
    \frac{2r_{\sss +} \lf(r_u^{3/2} - 2r_u^{1/2} \pm a\rt)
      \mp a \lf(r_u^2 \mp 2ar_u^{1/2} + a^2\rt)}
    {r_u^{3/4} \sqrt{r_u^{3/2} - 3r_u^{1/2} \pm 2a}}\\
    &= \frac{ \lf[ 2r_{\sss +}r_u^{3/2} - 4r_{\sss +}r_u^{1/2}
	+ 2\underbrace{a^2}_{r_{\sss +} r_{\sss -}} r_u^{1/2} \rt]
      \mp a
      \lf[ r_u^2 \underbrace{{}- 2r_{\sss +} + a^2}_{-r_{\sss +}^2}
      \rt]
    }
    {r_u^{3/4} \sqrt{r_u^{3/2} - 3r_u^{1/2} \pm 2a}}\\
    &= \frac{ 2r_{\sss +}r_u^{1/2} 
      \lf(r_u \underbrace{{}- 2 + r_{\sss -}}_{-r_{\sss +}} \rt)
     \mp a \lf(r_u - r_{\sss +}\rt) \lf(r_u + r_{\sss +}\rt)
     }
    {r_u^{3/4} \sqrt{r_u^{3/2} - 3r_u^{1/2} \pm 2a}}
    \elpunc{,}\\
    &= \frac{\lf(r_u - r_{\sss +}\rt)
      \lf[ 2r_{\sss +}r_u^{1/2} \mp a \lf(r_u + r_{\sss +}\rt) \rt]
    }
    {r_u^{3/4} \sqrt{r_u^{3/2} - 3r_u^{1/2} \pm 2a}}
  \end{split}
\end{align}
\end{widetext}
where we've used the facts that $r_{\sss \pm}$ are roots of $\Delta$
and that $r_{\sss +} r_{\sss -} = a^2$.  Analogously, the numerator of
the second factor in $C_5^{(t)}$ becomes
\begin{equation}
  2Er_{\sss -} - aL_z = \frac{\lf(r_u - r_{\sss -}\rt)
    \lf[ 2r_{\sss -}r_u^{1/2} \mp a \lf(r_u + r_{\sss -}\rt) \rt]
  }
  {r_u^{3/4} \sqrt{r_u^{3/2} - 3r_u^{1/2} \pm 2a}}
  \qquad ,
\end{equation}
leaving the coefficients $C_{4,5}^{(t)}$ as
\begin{align}
  \begin{split}
    C_4^{(t)} &= \frac{-4 r_{\sss +}}{r_{\sss +} - r_{\sss -}}
    \\
    &\mrph{=}
      {}\times\frac{1}{\sqrt{r_u^{3/2}
	  \lf(r_u^{3/2} - 3r_u^{1/2} \pm 2a\rt)}}
    \\
    &\mrph{=}
      {}\times
      \frac{2r_{\sss +}r_u^{1/2} \mp a \lf(r_u + r_{\sss +}\rt)}
      {\sqrt{r_{\sss +} \lf(r_a - r_{\sss +}\rt)}}
  \end{split}\\
  \begin{split}
    C_5^{(t)} &= \frac{4 r_{\sss -}}{r_{\sss +} - r_{\sss -}}
    \qquad\qquad\qquad\qquad\elpunc{.}\\
    &\mrph{=}
      {}\times\frac{1}{\sqrt{r_u^{3/2}
	  \lf(r_u^{3/2} - 3r_u^{1/2} \pm 2a\rt)}}
    \\
    &\mrph{=}
      {}\times
      \frac{2r_{\sss -}r_u^{1/2} \mp a \lf(r_u + r_{\sss -}\rt)}
      {\sqrt{r_{\sss -} \lf(r_a - r_{\sss -}\rt)}}
  \end{split}
\end{align}

For what follows, it will be useful to look at the signs of the
numerators
\begin{alignat}{2}
  &2r_{\sss +}r_u^{1/2} \mp a \lf(r_u + r_{\sss +}\rt) \, ,
  &&\quad\text{for $C_4$}
  \label{eq:C4numerator}\\
  &2r_{\sss -}r_u^{1/2} \mp a \lf(r_u + r_{\sss -}\rt) \, ,
  &&\quad\text{for $C_5$}
  \elpunc{.}
  \label{eq:C5numerator}
\end{alignat}
In the retrograde case (bottom sign), each is the sum of two
non-negative terms and thus strictly non-negative.  In the prograde
case (top sign), we can whether the sign depends on the values of
$r_u$ and $a$ by treating each of (\ref{eq:C4numerator}) and
(\ref{eq:C5numerator}) as quadratic function of the variable $y \equiv
r_u^{1/2}$.  Specifically, those functions will be \emph{negative}
when
\begin{alignat}{2}
  &a y^2 - 2 r_{\sss +} y + a r_{\sss +} > 0 \, ,
  &&\quad\text{for $C_4$} 
  \label{eq:C4numeratorquady}\\
  &a y^2 - 2 r_{\sss -} y + a r_{\sss -} > 0 \, ,
  &&\quad\text{for $C_5$}
  \elpunc{.}
  \label{eq:C5numeratorquady}
\end{alignat}
Since the expressions above have positive quadratic coefficients, the
inequalities (\ref{eq:C4numeratorquady}) and
(\ref{eq:C5numeratorquady}) are satisfied when
\begin{equation}
 y < \frac{r_{\sss +}}{a} \lf( 1 - \sqrt{1 - r_{\sss -}} \rt)
 \quad\text{or}\quad
 y > \frac{r_{\sss +}}{a} \lf( 1 + \sqrt{1 - r_{\sss -}} \rt)
  \, ,
  \label{eq:C4numeratorineqsy}
\end{equation}
for $C_4$ and
\begin{equation}
y < \frac{r_{\sss -}}{a} \lf( 1 - \sqrt{1 - r_{\sss +}} \rt)
\quad\text{or}\quad
y > \frac{r_{\sss -}}{a} \lf( 1 + \sqrt{1 - r_{\sss +}} \rt)
  \, ,
  \label{eq:C5numeratorineqsy}
\end{equation}
for $C_5$.
In the case of \eqref{eq:C5numeratorineqsy}, the radicand $1 - r_{\sss
+} = -\sqrt{1 - a^2}$ is strictly negative\footnote{For $a = 1$, the
radicand is 0, not negative.  However, in this scenario, $r_{\sss -} =
1$ and the quadratic expression in (\ref{eq:C5numeratorquady}) has a
double root at $y = 1 \implies r_u = 1$.  Since $r_u \geq 1$ for all
$a$, then even in the $a = 1$ case, the quadratic in
(\ref{eq:C5numeratorquady}) will be non-negative.  Of course, the
$a=1$ case for any analysis of orbital motion must be handled
carefully since the $r$ coordinate values of the inner and outer
horizons, the itco, the ibco and the isco are all unphysically
degenerate in the maximal spin case.} and the roots are complex.  The
quadratic expression is therefore always positive, and
(\ref{eq:C5numerator}) is always negative.

For (\ref{eq:C4numeratorineqsy}), we note that since
\begin{alignat}{2}
  &\frac{1 - \sqrt{1 - r_{\sss -}}}{a} < 1
  &&\quad\text{for $0 < a < 1$}
  \quad ,
\end{alignat}
the lower root is subhorizon and thus irrelevant (because $r_u >
r_{\sss +}$).  So what we must check is whether we can ever have
\begin{equation}
  r_u = y^2 > \frac{r_{\sss +}^2}{a^2}
  \lf( 1 - \sqrt{1 - r_{\sss -}} \rt)^2
  \quad .
\label{eq:C4numeratortest}
\end{equation}
In fact, \eqref{eq:C4numeratortest} is never satisfied for prograde
orbits.  To see why, recall \cite{Bardeen1972} that for prograde
equatorial orbits,
\begin{align}
  \begin{split}
    r_{\tm{isco}} &= 3 + Z_2
    -\lf[
      \lf(3 - Z_1\rt) \lf(3 + Z_1 + 2Z_2\rt)
      \rt]^{1/2}\\
    Z_1 &= 1 + \lf(1 - a^2\rt)^{1/3}
    \lf[
      \lf(1 + a\rt)^{1/3} + \lf(1 - a\rt)^{1/3}
      \rt]
    \quad .\\
    Z_2 &= \lf(3a^2 + Z_1^2\rt)^{1/2}
  \end{split}
\label{eq:ribcopro}
\end{align}
A simple plot (not included here) shows that
\begin{alignat}{2}
  &r_{\tm{isco}} < \frac{r_{\sss +}^2}{a^2}
  \lf( 1 - \sqrt{1 - r_{\sss -}} \rt)^2
  &&\quad \text{for $0 < a < 1$}
  \quad .
\label{eq:ribcovsrcrit}
\end{alignat}
Since $r_u < r_{\tm{isco}}$ for all (eccentric) homoclinic orbits,
(\ref{eq:C4numeratortest}) is never satisfied, and
(\ref{eq:C4numerator}) is always positive.

The upshot is that we can write the $C_{\sss 4,5}^{(t)}$'s so that
every factor outside a radical is positive.  Specifically,
\begin{widetext}
\begin{align}
  \begin{split}
    C_4^{(t)} &= 
    -\frac{4 r_{\sss +}}{r_{\sss +} - r_{\sss -}}
    \times \sqrt{1 - E^2}
      {}\times
      \frac{2r_{\sss +}r_u^{1/2} \mp a \lf(r_u + r_{\sss +}\rt)}
	   {\sqrt{r_{\sss +}
	       \lf[ 2 - \lf(2r_u + r_{\sss +}\rt)\lf(1 - E^2\rt)
		 \rt]
	       r_u^{3/2}\lf(r_u^{3/2} - 3r_u^{1/2} \pm 2a\rt)
	     }
	   }
  \end{split}
  \\
  \begin{split}
    C_5^{(t)} &= 
    -\frac{4 r_{\sss -}}{r_{\sss +} - r_{\sss -}}
    \times \sqrt{1 - E^2}
      {}\times
      \frac{-2r_{\sss -}r_u^{1/2} \pm a \lf(r_u + r_{\sss -}\rt)}
	   {\sqrt{r_{\sss -}
	       \lf[ 2 - \lf(2r_u + r_{\sss -}\rt)\lf(1 - E^2\rt)
		 \rt]
	       r_u^{3/2}\lf(r_u^{3/2} - 3r_u^{1/2} \pm 2a\rt)
	     }
	   }
  \end{split}
  \quad ,
\end{align}
\end{widetext}
where we have used equation \eqref{eq:2ruplusra} to rewrite the factors
$(r_a - r_{\sss \pm})$ in the radicands of the denominators as
\begin{equation}
  r_a - r_{\sss \pm} = \frac{2 - \lf(2r_u + r_{\sss \pm}\rt)
  \lf(1 - E^2\rt)
  }
  {1 - E^2} \quad .
\end{equation}
Since the numerators of the factors on the second lines are now
manifestly positive, they can be brought under the radical sign
without having to worry about stray factors of $-1$.  We are left with
\begin{widetext}
\begin{align}
  \begin{split}
    C_4^{(t)} &= 
    -\frac{4 r_{\sss +}}{r_{\sss +} - r_{\sss -}}
    \times \sqrt{1 - E^2}
      {}\times
      \sqrt{
	\frac{
	  \lf[
	    2r_{\sss +}r_u^{1/2} \mp a \lf(r_u + r_{\sss +}\rt)
	  \rt]^2
	}
	     {	
	       r_{\sss +}
	       \lf[ 2 - \lf(2r_u + r_{\sss +}\rt)\lf(1 - E^2\rt)
		 \rt]
	       r_u^{3/2}\lf(r_u^{3/2} - 3r_u^{1/2} \pm 2a\rt)
	     }
	   }
  \end{split}
  \label{eq:C4messyfinal}\\
  \begin{split}
    C_5^{(t)} &= 
    -\frac{4 r_{\sss -}}{r_{\sss +} - r_{\sss -}}
    \times \sqrt{1 - E^2}
      {}\times
      \sqrt{
	\frac{
	  \lf[
	    -2r_{\sss -}r_u^{1/2} \pm a \lf(r_u + r_{\sss -}\rt)
	  \rt]^2
	}
	     {	
	       r_{\sss -}
	       \lf[ 2 - \lf(2r_u + r_{\sss -}\rt)\lf(1 - E^2\rt)
		 \rt]
	       r_u^{3/2}\lf(r_u^{3/2} - 3r_u^{1/2} \pm 2a\rt)
	     }
	   }
  \end{split}
  \quad ,
  \label{eq:C5messyfinal}
\end{align}
\end{widetext}
Finally, each of the large radicands in (\ref{eq:C4messyfinal}),
(\ref{eq:C5messyfinal}) is 1.  To see this, we use equation
(\ref{app:eq:Ecirc}) to rewrite the $1 - E^2$ in each denominator as
\begin{equation}
  1 - E^2 = \frac{r_u^2 - 4r_u \pm 4ar_u^{1/2} - a^2}
  {r_u^{3/2}\lf(r_u^{3/2} - 3r_u^{1/2} \pm 2a\rt)}
\label{eq:1minusEsq}
\end{equation}
so that distributing the $r_u^{3/2}\lf(r_u^{3/2} - 3r_u^{1/2} \pm
2a\rt)$ in the denominators of the radicands leaves them in the form
\begin{alignat}{2}
\begin{split}
  &r_{\sss +} \lf[
    2 r_u^{3/2}\lf(r_u^{3/2} - 3r_u^{1/2} \pm 2a\rt)
    \rt.\\
    &\quad\lf.
       {}- \lf( 2r_u + r_{\sss +} \rt)
       \lf( r_u^2 - 4r_u \pm 4ar_u^{1/2} - a^2 \rt)
       \rt]
  \,,
\end{split}
&&\quad \text{for $C_4^{(t)}$}
\label{eq:C4denomradicand}\\
\begin{split}
  &r_{\sss -} \lf[
    2 r_u^{3/2}\lf(r_u^{3/2} - 3r_u^{1/2} \pm 2a\rt)
    \rt.\\
    &\quad\lf.
    {}- \lf( 2r_u + r_{\sss -} \rt)
    \lf( r_u^2 - 4r_u \pm 4ar_u^{1/2} - a^2 \rt)
    \rt]
\,,
\end{split}
  &&\quad \text{for $C_5^{(t)}$}
\label{eq:C5denomradicand}
\end{alignat}
Multiplying out the numerators and denominators and grouping them by
powers of $r_u$ then shows that they are identical, for both prograde
and retrograde orbits.

The final expressions for the coefficients $C_{\sss 4,5}^{(t)}$ are
compact.  Noting that $r_{\sss +} - r_{\sss -} = 2 \sqrt{1 - a^2}$,
those expressions are
\begin{align}
  C_4^{(t)} &= -\frac{2 r_{\sss +}}{\sqrt{1 - a^2}}
    \times \sqrt{1 - E^2}
    \label{eq:C4tfinal}\\
  C_5^{(t)} &= -\frac{2 r_{\sss -}}{\sqrt{1 - a^2}}
    \times \sqrt{1 - E^2}
    \label{eq:C5tfinal}
    \elpunc{.}
\end{align}
To get the corresponding $\varphi$ coefficients, note that
\begin{equation}
  \begin{split}
    2aE - L_zr_{\sss \mp} &= \frac{1}{r_{\sss \pm}}
    \lf[
      2aEr_{\sss \pm} - L_z\underbrace{r_{\sss \mp} r_{\sss \pm}}_{a^2}
      \rt]
    \\
    &= \frac{a}{r_{\sss \pm}}
    \lf[
      2Er_{\sss \pm} - aL_z
      \rt]
  \end{split}
  \quad .
\end{equation}
Thus,
\begin{align}
  C_4^{(\varphi)} &= \frac{1}{2} \frac{a}{r_{\sss +}} C_4^{(t)}
  =  -\frac{a}{\sqrt{1 - a^2}} \times \sqrt{1 - E^2}
  \label{eq:C4phifinal}\\
  C_5^{(\varphi)} &= \frac{1}{2} \frac{a}{r_{\sss -}} C_5^{(t)}
  =  -\frac{a}{\sqrt{1 - a^2}} \times \sqrt{1 - E^2}
  \label{eq:C5phifinal}
  \elpunc{.}
\end{align}

To summarize, once simplified, the coefficients in (\ref{eq:Cs})
become
\begin{alignat}{2}
  C_1^{(\tau)} &= 1 \qquad\,\,
  C_2^{(\tau)} = \frac{2}{1 - E^2}\quad
  & C_3^{(\tau)} &= \frac{2}{\gamma\lambda_r}
  \sqrt{1 - E^2}
  \nonumber\\
  C_1^{(t)} &= E
  & C_2^{(t)} &= 2E\lf(\frac{3 - 2E^2}{1 - E^2}\rt)
  \nonumber\\
  C_3^{(t)} &= \frac{2}{\lambda_r} \sqrt{1 - E^2}
  &
  C_3^{(\varphi)} &= 2 \frac{\Omega_u}{\lambda_r} \sqrt{1 - E^2}
  \qquad\quad .
  \label{eq:Csfinal}\\
  C_4^{(t)} &= -\frac{2 r_{\sss +}}{\sqrt{1 - a^2}} \sqrt{1 - E^2}
  &
  C_4^{(\varphi)} &= -\frac{a}{\sqrt{1 - a^2}} \sqrt{1 - E^2}
  \nonumber\\
  C_5^{(t)} &= -\frac{2 r_{\sss -}}{\sqrt{1 - a^2}} \sqrt{1 - E^2}
  &
  C_5^{(\varphi)} &= -\frac{a}{\sqrt{1 - a^2}} \sqrt{1 - E^2}
  \nonumber
\end{alignat}

\subsection{Analytic expressions for homoclinic orbits}
We can now put everything together from the prior subsections.
Looking back at equations \eqref{eq:tauhomoCfs} -
\eqref{eq:phihomoCfs} and substituting from (\ref{eq:fs}) and
(\ref{eq:Csfinal}), we arrive at the final expressions for all the
dynamical variables:
\begin{widetext}
\begin{align}
  \begin{split}
    \tau(r) &= \frac{1}{\sqrt{1 - E^2}}\sqrt{r(r_a - r)}
    + \frac{2}{\lf(1 - E^2\rt)^{3/2}}
    \atan\sqrt{\frac{r_a - r}{r}}
    + \frac{2}{\gamma\lambda_r}
    \atanh\sqrt{\frac{r_u}{r_a - r_u}\frac{r_a - r}{r}}    
  \end{split}\\
  \begin{split}
    t(r) &= \frac{E}{\sqrt{1 - E^2}}\sqrt{r(r_a - r)}
    + 2E\frac{\lf(3 - 2E^2\rt)}{\lf(1 - E^2\rt)^{3/2}}
    \atan\sqrt{\frac{r_a - r}{r}}
    + \frac{2}{\lambda_r}
    \atanh\sqrt{\frac{r_u}{r_a - r_u}\frac{r_a - r}{r}}
    \\
    &\mrph{=} {}- \frac{2 r_{\sss +}}{\sqrt{1 - a^2}}
    \atanh
    \sqrt{
      \frac{r_{\sss +}}{r_a - r_{\sss +}}
      \frac{r_a - r}{r}
    } 
    - \frac{2 r_{\sss -}}{\sqrt{1 - a^2}}
    \atanh
    \sqrt{
      \frac{r_{\sss -}}{r_a - r_{\sss -}}
      \frac{r_a - r}{r}
    }
  \end{split}
  \quad .\\
  \begin{split}
    \varphi(r) &= 2 \frac{\Omega_u}{\lambda_r}
    \atanh\sqrt{\frac{r_u}{r_a - r_u}\frac{r_a - r}{r}}
    - \frac{a}{\sqrt{1 - a^2}}
    \atanh
    \sqrt{
      \frac{r_{\sss +}}{r_a - r_{\sss +}}
      \frac{r_a - r}{r}
    }
    - \frac{a}{\sqrt{1 - a^2}}
    \atanh
    \sqrt{
      \frac{r_{\sss -}}{r_a - r_{\sss -}}
      \frac{r_a - r}{r}
    }
  \end{split}
\end{align}
\end{widetext}

\bibliographystyle{aip.bst}
\bibliography{hcone}

\end{document}